\documentclass[aps,prd,nofootinbib,reprint,preprintnumbers,superscriptaddress]{revtex4-1}
\UseRawInputEncoding
\usepackage{supertabular}
\usepackage{lipsum}
\usepackage{blindtext}
\usepackage{amsfonts}
\usepackage{tensor}
\usepackage{graphicx}
\usepackage{pict2e}
\usepackage{balance}
\usepackage{blindtext}

\usepackage{amssymb, amsmath,environ}

\usepackage{color}
\usepackage[colorlinks=true,urlcolor=black,linkcolor=black,citecolor=red]{hyperref}


\newcommand{\bse}{\begin{subequations}}
	\newcommand{\ese}{\end{subequations}}
\newcommand{\be}{\begin{equation}}
\newcommand{\ee}{\end{equation}}

\NewEnviron{eqsplit}{%
	\begin{equation}\begin{split}
	\BODY
	\end{split}\end{equation}
}

\makeatletter
\newcommand*\bigcdot{\mathpalette\bigcdot@{.5}}
\newcommand*\bigcdot@[2]{\mathbin{\vcenter{\hbox{\scalebox{#2}{$\m@th#1\bullet$}}}}}
\makeatother

\newcommand{\bea}{\begin{eqnarray}}
\newcommand{\eea}{\end{eqnarray}}
\newcommand{\ba}{\begin{array}}
	\newcommand{\ea}{\end{array}}

\newcommand{\R}{\mathcal{R}}

\newcommand{\la}{\langle}
\newcommand{\ra}{\rangle}

\begin{document}
\preprint{MITP-23-004}		
	
	\title{Flow Distribution Analysis as A Probe of Nuclei Deformity}

	%
	%
	%
	%
	%
	%
	%
	%
	
	\author{Hadi Mehrabpour}
	\email[]{mehrabph@uni-mainz.de}
	\affiliation{PRISMA$^{+}$ Cluster of Excellence \& Mainz Institute for Theoretical Physics,
		Johannes Gutenberg-Universit\"at Mainz, 55099 Mainz, Germany}
	
	\author{S.~M.~A.~Tabatabaee}
	\email[]{tabatabaee@ipm.ir}
	\affiliation{School of Particles and Accelerators, Institute for Research in Fundamental Sciences (IPM), P.O. Box 19395-5531, Tehran, Iran}

\begin{abstract}
We study the flow harmonic distribution in deformed nuclei. To do this, we use the standard Gram-Charlier method to find the higher-order correction to the well-known Bessel-Gaussian distribution. We find that, apart from the necessity of including a shift parameter $\bar v_n$, the modified flow distribution describes the flow distribution of quadrupole and octupole deformation accurately. Using the shifted radial distribution, arising from this method, we scrutinize the effect of deformation on flow distribution. Assuming a linear relation between observables of spherical and deformed collisions, $\mathcal{O}_D=\mathcal{O}_S+\left(\sum_{m=2}a_{k,m} \beta_m\right)^{2k}$, for events with a fixed centrality, we compare the flow distribution of deformed and spherical nuclei. We also propose a way to measure $\bar{v}_2$ in asymmetric nuclei collisions.     
\end{abstract}

\maketitle
\section{Introduction}\label{sec1} 
Quantum Chromodynamic (QCD), the theory of strong interaction, undergoes certain phase transition at high temperature to a plasma of quarks and gluons. Heavy-ion collision (HIC) at Relativistic Heavy Ion Collider (RHIC) and Large Hadron Collider (LHC) provide this opportunity to study the phase structure of QCD and the properties of quark-gluon plasma (QGP) \cite{Ackermann:2000tr,Lacey:2001va,Park:2001gm,Aamodt:2010pa,ALICE:2011ab,Chatrchyan:2012ta,ATLAS:2011ah,Aad:2014vba}. One feature of this plasma is collective behavior which can be successfully described by the hydrodynamic models \cite{Luzum:2008cw}. Among the various probes to study the dynamics of relativistic HIC, the most used one, is the anisotropic flow that is quantified with harmonics $v_n$, measuring the azimuthal asymmetry of the emitted hadrons. The flow distribution and the cumulants are used to gain more information on the even-by-event fluctuations \cite{Ollitrault:1992bk}. This insight shed light on the collision geometry, quantum fluctuations at initial state as well as the effects of different evolution stages in heavy-ion process \cite{Schenke:2012wb,Miller:2003kd}. From experimental point of view, the distribution of $v_{2}$ and $v_{3}$ are accessible through the unfolding method \cite{Jia:2013tja}. This leads to an observation of a Bessel-Gaussian distribution for collision of spherical nuclei, i.e., Pb-Pb, in the central collision \cite{Aad:2013xma}. On the other hand, cumulants can be obtained as a measure of multi-particle correlation functions \cite{Borghini:2001vi}. 
Measuring the correlation of particles gives this chance to map the shape of nuclei \cite{Bally:2022vgo}. 

Conventionally, in low-energy nuclear physics, a Woods-Saxon profile describes the density of nucleons inside a nucleus,
\begin{equation}
\rho(r,\theta,\phi)\propto\frac{1}{1+e^{\frac{r-R(\theta,\phi)}{a_0}}},
\end{equation}
where $a_0$ and $R(\theta,\phi)$ are the surface diffuseness and the nuclear surface parameter, respectively.
In general, to take into account the deformation of the nucleus, $R(\theta,\phi)$ is expanded in terms of spherical harmonics $Y_{\ell}^{m}(\theta,\phi)$. In particular the quadrupole deformation is defined by  $R(\theta,\phi)=R_0(1+\beta_2(\cos\gamma Y_2^0(\theta,\phi)+\sin\gamma Y_2^2(\theta,\phi)))$ \cite{Moller:1994cnm}.
Here, $R_0$ is the half-density radius, $\gamma$ determines the relative length of the three axes of the ellipsoid, and $\beta_2$ is the magnitude of quadrupole deformation.
Experimental results show a noticeable difference between the $v_{2}$ of deformed and spherical nuclei \cite{STAR:2015mki}, with the largest difference being in the most central collisions. Besides the quadrupole deformation, axial deformation may also significantly affect observables. This form of deformation arises due to the breaking of parity symmetry in the intrinsic nuclear shape. This kind of deformity is modeled by including $\beta_3Y_3^0(\theta,\phi)$ in the nuclear surface parameter. 

\par
The structure of this paper is as follows: In Sec.~\ref{sec2}, we present a brief review of standard Gram-Charlier series method. Using this approach, we find the cumulants of flow harmonic in Sec.~\ref{sec3}. We argue that to include the effect of deformation on the cumulants we have to consider a shift parameter in the definition. Then we obtain the flow distribution for the magnitude of flow harmonics. We observe, the conventional Bessel-Gaussian is not appropriate for the central collisions. Once, we consider higher order correction the data is explained accurately. Having the appropriate distribution, we compare the spherical and deformed nuclei in Sec.~\ref{sec5}. We see, for particular case of quadrupole deformation, the corresponding distribution is broader than the spherical one. In Sec.~\ref{sec6}, we present an approach to observe the shift parameter in experiments. We summarize in Sec.~\ref{sec.con} and present our concluding remarks.
  	
\section{Method of Analysis}\label{sec2}
In this section, we use the so-called standard Gram-Charlier (sGC) series to find the distribution of any random variable. In case where the related information of the desired variable, such as its moments, is incompatible with Gaussian distributions, this series can be used to modify Gaussian distributions. This method relates the probability distribution $P(\mathbf{Z})$ to the Gaussian distribution by applying an appropriate differential operator \cite{Kendall:1945,Cramer:1999,Krzanowski:2000}. To do this, let us start with the characteristic function $\Phi_{\mathbf{Z}}(\mathbf{t})$ of a k-dimensional random vector $\mathbf{Z}$, which is defined as:
\begin{equation}\label{qq1} 
\Phi_{\mathbf{Z}}(\mathbf{t})=\int \mathbf{d}\mathbf{Z}\; e^{i\mathbf{t}^{T}\mathbf{Z}} P(\mathbf{Z}).
\end{equation}
Here, $\mathbf{t}$ belongs to the $\mathcal{C}^k$ ($\R^k$) when $\mathbf{Z}$ is complex (real). As it turns out, if the random vector $\mathbf{Z}$ has a moment-generating function $\mathcal{M}$, the domain of the characteristic function can be extended to the complex plane, and thus we have $\Phi(−i\mathbf{t})=\mathcal{M}(\mathbf{t})$. Moreover, an alternative definition of the second characteristic function $\mathcal{M}(t)$ is given by $\mathcal{K}(\mathbf{t})=\log \mathcal{M}(\mathbf{t})$. Thus, this leads to the cumulants of random vector as $\mathcal{K}^{(n)}(\mathbf{t})|_{t=0}=(\log \mathcal{M}(\mathbf{t}))^{(n)}|_{\mathbf{t}=0}$\footnote{Assuming a real random variable $x$ the corresponding equation leads to:
	\begin{align*}
	1+\sum_{n=1}^{\infty}\frac{\mu_n t^n}{n!}=\exp\Big(\sum_{n=1}^{\infty}\frac{\kappa_n t^n}{n!} \Big).
	\end{align*}
	where for a particular choice $\mu=\la x\ra$, it implies:
	\begin{align*}
	\mu_1&=\mu =k_1,\\
	\mu_2&=\mu ^2+\sigma ^2=k_1{}^2+k_2, \\
	\mu_3&=\mu ^3+3 \mu  \sigma ^2+k_3=k_1{}^3+3k_1k_2+k_3.
	\end{align*}
}.
Furthermore, the probability function defined in Eq.~\ref{qq1} is determined by an inverse Fourier transformation \cite{Cramer:1999} as following:
\begin{equation}\label{qq2} 
P(\mathbf{Z})=\frac{1}{2\pi}\int \mathbf{d}\mathbf{t}\; e^{-i\mathbf{t}^T\mathbf{Z}} \Phi_{\mathbf{Z}}(\mathbf{t}).
\end{equation}   
Expanding the characteristic function up to second order, writing $(i\mathbf{Z})^n$ in terms of an appropriate differential operator, we arrive at an expression for $P(\mathbf{Z})$. Following the steps described above to find the distribution function for a real random variable $x$, we find the corrections to the conventional Gaussian distribution. The corresponding terms are given in terms of the probabilist's Hermite polynomials, $He_n$, as	
$\sum_{n=3}^{\infty}\frac{\kappa_n}{n!\sigma^n}He_{n}(\frac{x-\mu}{\sigma})$. The sGC method aims to find the non-Gaussianity correction to obtain a complete description of our variable. This leads us to focus on the characteristic function instead. Since this quantity gives us the desired information about the moments and cumulants, it provides insight into the probability distribution. It is known that both collision geometry and event-by-event fluctuations are encoded in flow harmonic distributions $P(v_n)$ as well as cumulants \cite{Voloshin:2007pc}. In the following sections, we study the connection between them to gain a deeper insight on the effects of nucleus deformation using sGC series in the two cases of symmetric and deformed ion collisions.        

\section{spherical and deformed nuclei collisions}\label{sec3}
In this section, we examine the effect of deformation on the flow anisotropy. To start, we present the $2k$-particle correlation functions $c_n\{2k\}$ \cite{Borghini:2001vi} as well as the shifted cumulants \cite{Mehrabpour:2020wlu} for symmetric and deformed ions. The present work aims to study the effects of deformation on cumulants, resulting from the initial stage of collision of nuclei. To do this, we use the approximate relation between $v_n$ and the initial anisotropy $\varepsilon_n$ for second and third harmonics: $v_n=\alpha_n\varepsilon_n$ \cite{Giacalone:2017dud,Giacalone:2018apa}. We do not need to compute $v_2$ and $v_3$ by means of full hydrodynamic simulations. The $\alpha_n$ is a response coefficient that depends on the properties of the medium, such as its viscosity, and it is the same for all events at a given centrality\footnote{Since this linear response works for $n=2,3$ not higher harmonics and gives us a simple picture of the relationship between initial and final states, we do not discuss higher harmonics here. The method employed here, however, works to study higher flow harmonic as well while that would be complicated.}. Its value has been determined at both RHIC and LHC energies (see Ref.\cite{Giacalone:2019pca}). Thus, we have generated data for PbPb and UU, as well as ZrZr, collisions at the center-of-mass energy $\sqrt{s_{NN}}=5.02\;\text{TeV}$ and $200\;\text{GeV}$, respectively, motivated by LHC \cite{ALICE:2022xir} and RHIC \cite{STAR:2015mki} experiments \footnote{We performed a same analysis for UU collisions at $5.02\;\text{TeV}$ center-of-mass energies and the results are exactly the same.}. Also, we use the same T$_{\text{R}}$ENTO parametrization
\footnote{In this study, we use the geometric thickness function with $p=0$. The nucleus thickness function is a superposition of the nucleon thickness function whose Gaussian width is chosen as $w=0.5$. Moreover, the fluctuation of nucleon thickness function is considered by a gamma distribution with variance $1/k$ where we have used $k=1$.} for these simulations at different centrality classes to have the same situation for both deformed and symmetric nuclei. To better understand the effect of deformation in deformed collisions, we consider three sets of deformation parameters for UU collisions as follows:
\begin{equation*}
\begin{split}
1)&\; \beta_2=0 \;\text{and}\; \beta_3=0 \;\text{: spherical U},\\
2)&\; \beta_2=0.265 \;\text{and}\; \beta_3=0 \;\text{: effect of $\beta_2$ },\\
3)&\; \beta_2=0.265 \;\text{and}\; \beta_3=0.1 \;\text{: effect of $\beta_2$ and $\beta_3$}.
\end{split}	
\end{equation*}
\begin{figure}[t!]
	\begin{tabular}{c}
		\includegraphics[scale=0.4]{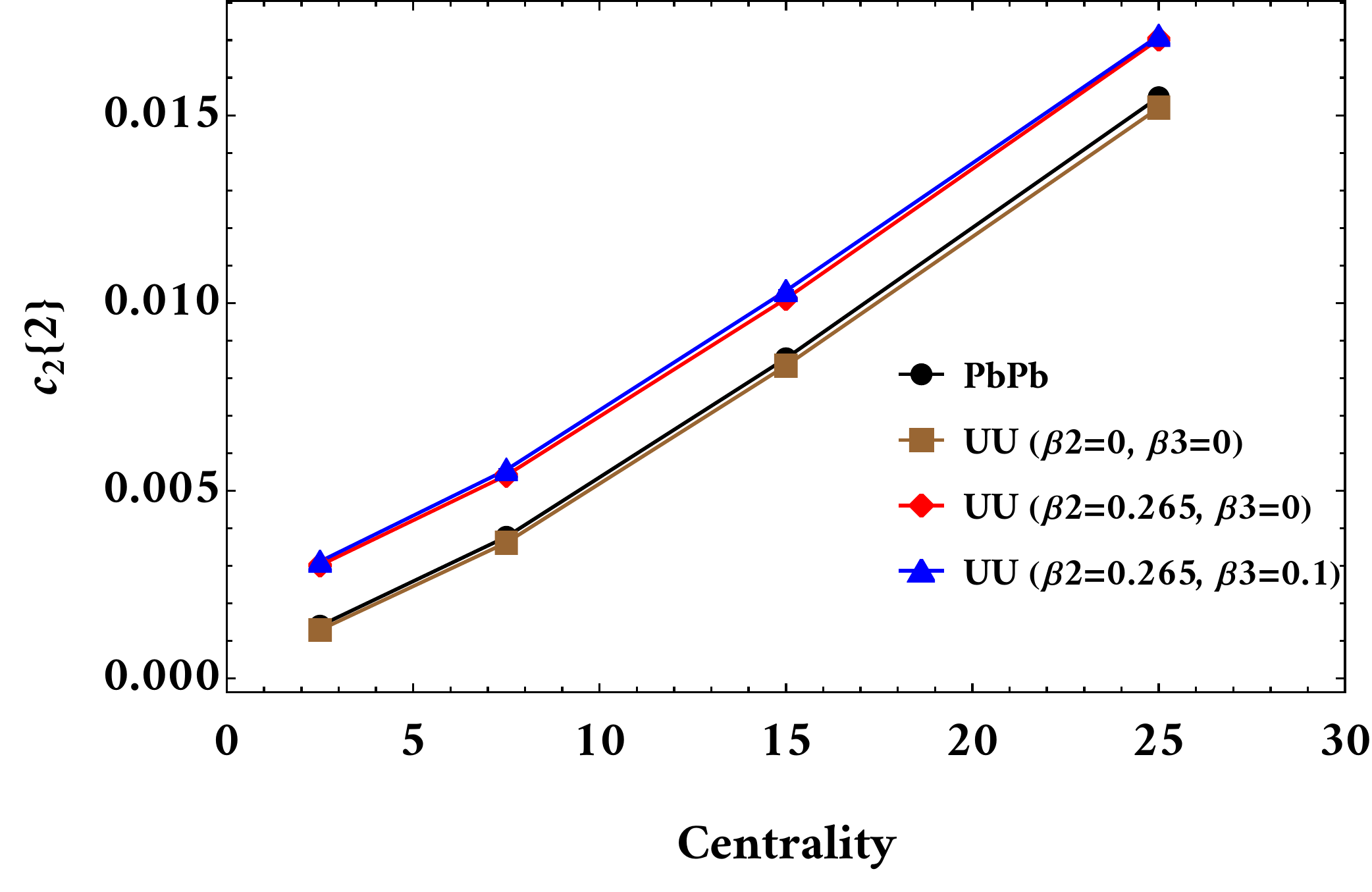}\\
		\includegraphics[scale=0.4]{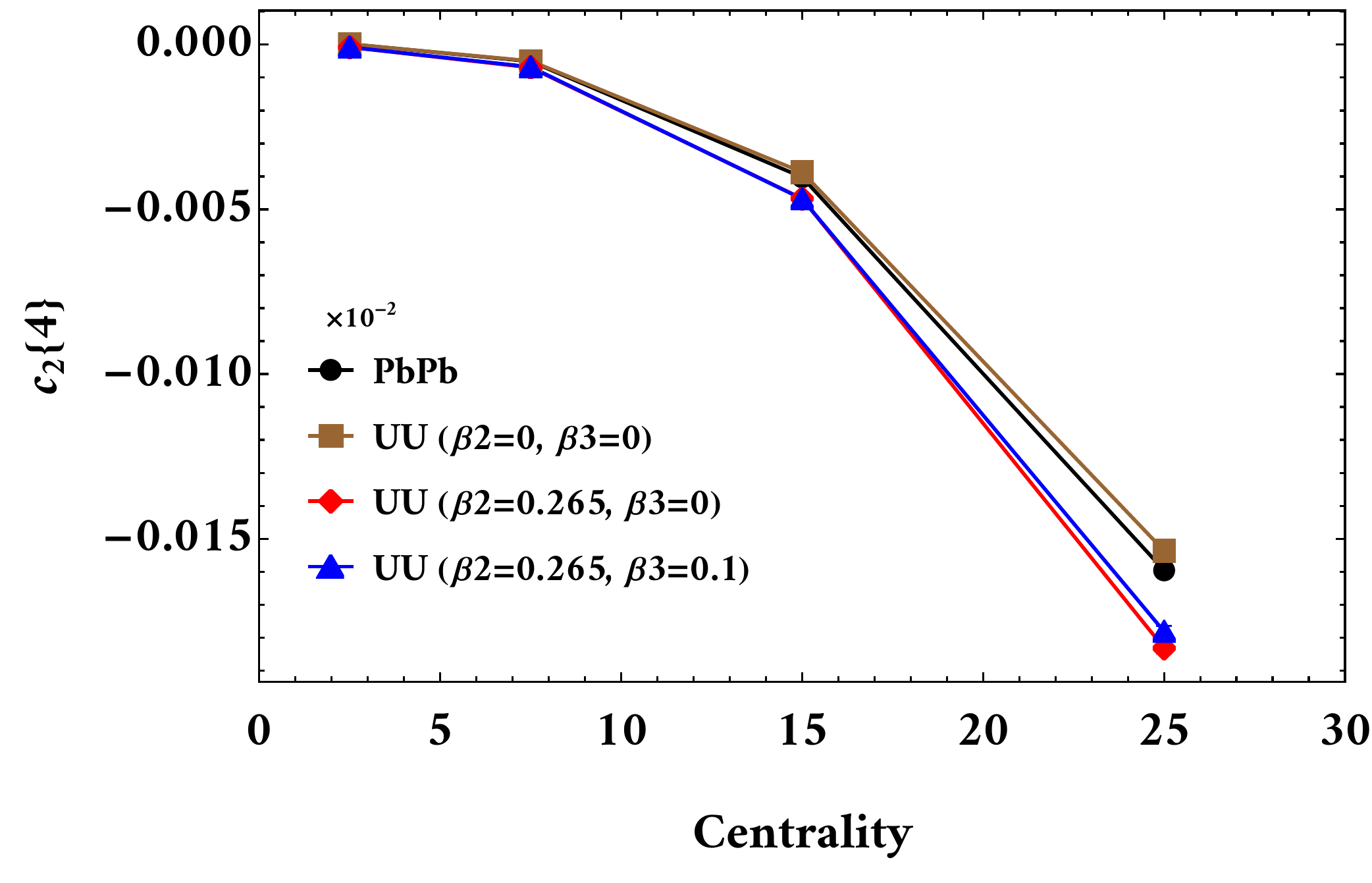}\\
		\includegraphics[scale=0.4]{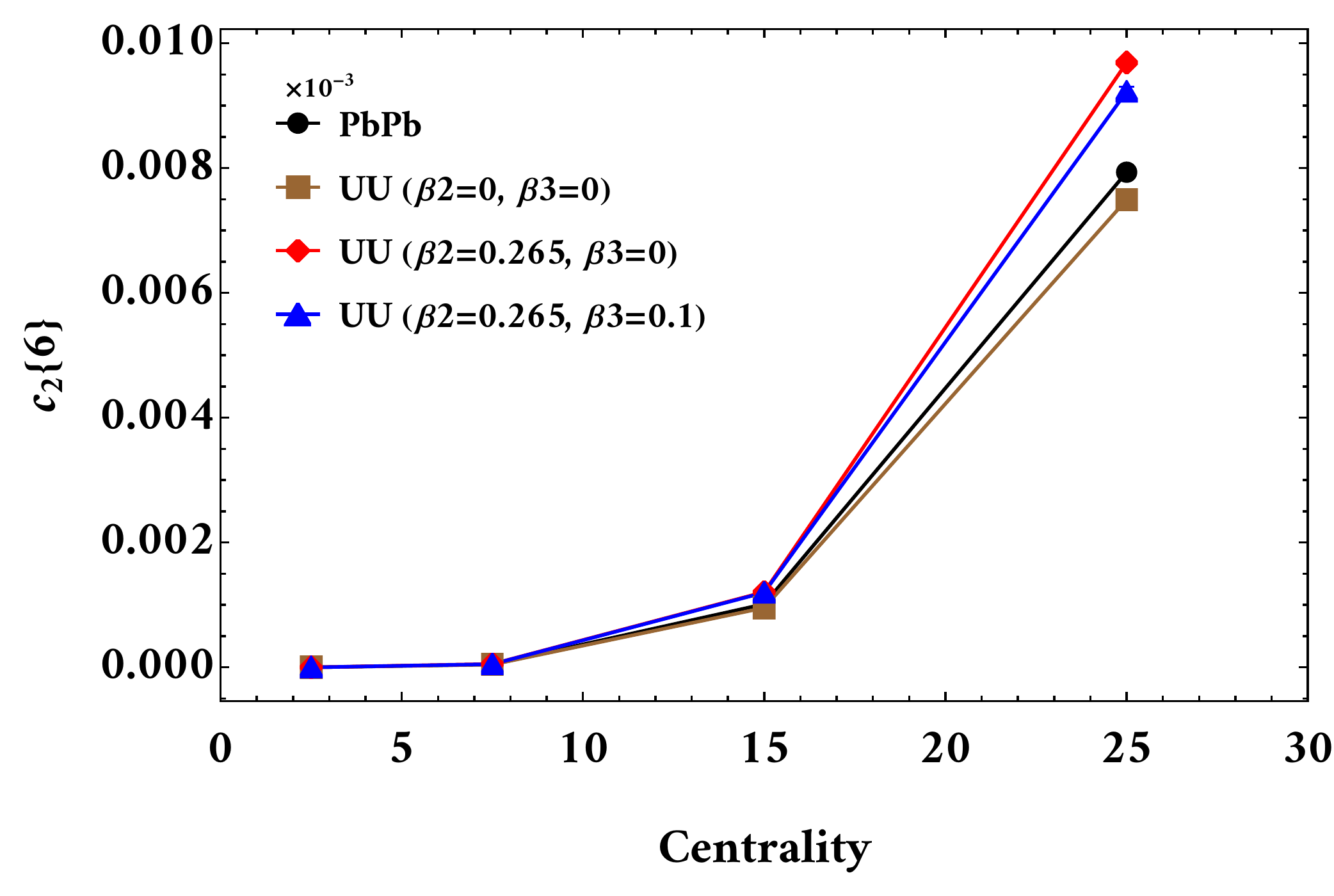}
	\end{tabular}		
	\caption{(Color online) Comparing $2k$-particle correlation functions $c_n\{2k\}$ for spherical and deformed nuclei collisions. Data has been generated at the center-of-mass energy $\sqrt{s_{NN}}=5.02\;\text{TeV}$.} 
	\label{fig1}
\end{figure}
In the previous section, it was shown that one way to study a random variable is to scrutinize its moments, cumulants, and probability distribution. We propose that, in order to obtain more insight into the impact of deformation, we need to study flow harmonics. Thus, we will compare the observables of symmetric and deformed nuclei. To show how this method works in the flow studies, we plug $t=\frac{1}{2}(t_x-it_y)$ and $Z=v_{x,n}+iv_{y,n}$ in Eq.~\ref{qq1}  for any harmonics (see more details in Ref.\cite{Mehrabpour:2020wlu}). Following these considerations, we arrive at the $2k$-particle correlation functions \cite{Borghini:2001vi}:
\begin{equation}
c_n\{2\}=\la v_n^2\ra,\; c_n\{4\}=\la v_n^4\ra-2\la v_n^2\ra^2,\;\cdots.
\end{equation}
In Fig.~\ref{fig1}, we plot the two particle correlation function $c_2\{2k\}$ for $k=1,2,3$ at various centralities.
As demonstrated in this figure, we find a finite difference between the correlation function $c_2\{2\}$ of symmetric and asymmetric nuclei, whose magnitude is larger for deformed ones. Concerning the four- and six-particle correlation function, we see there is no significant difference at $0-5\%$ and $5-10\%$ centralities, whereas we expected to observe a considerable deformation effect.  

\begin{figure}[t!]
	\begin{tabular}{c}
		\includegraphics[scale=0.2]{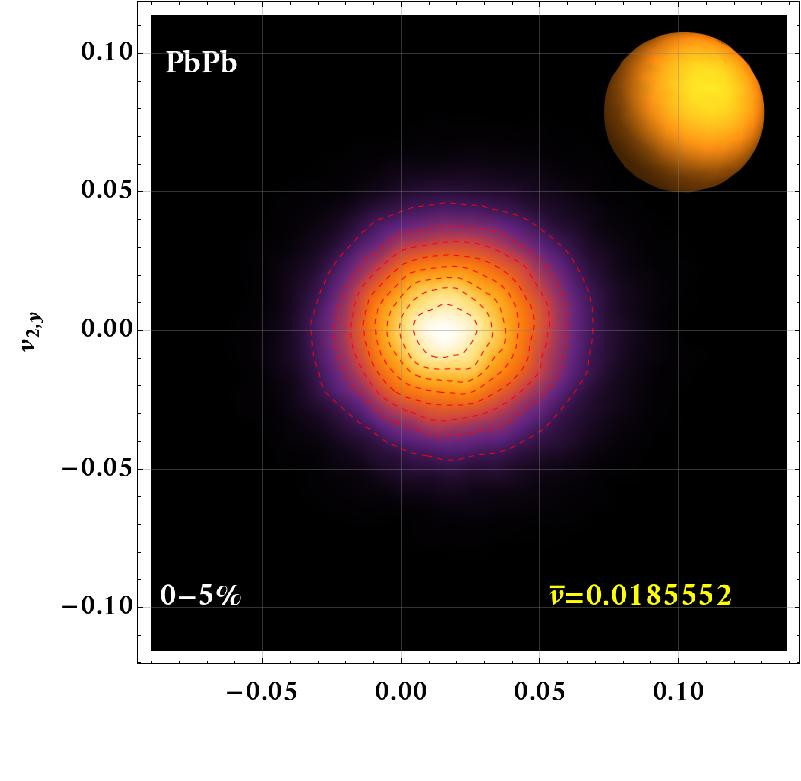}
		\includegraphics[scale=0.2]{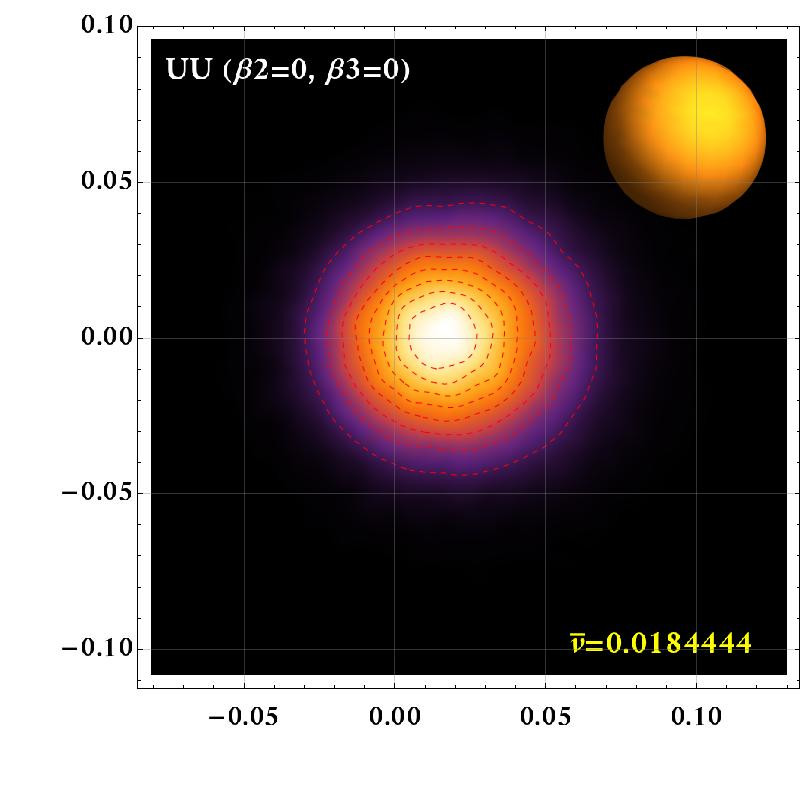}\\
		\includegraphics[scale=0.2]{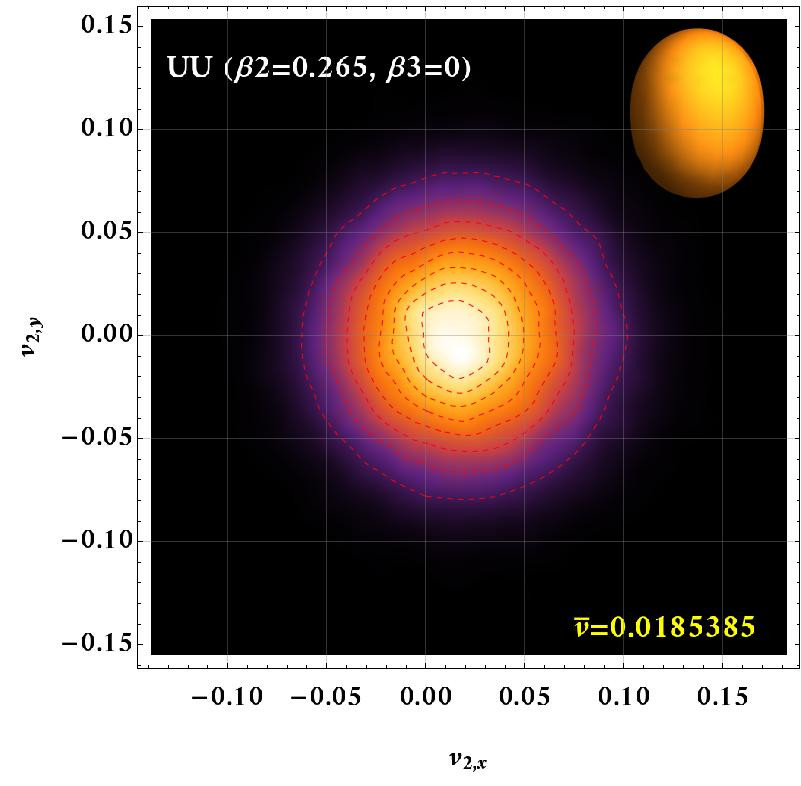}
		\includegraphics[scale=0.2]{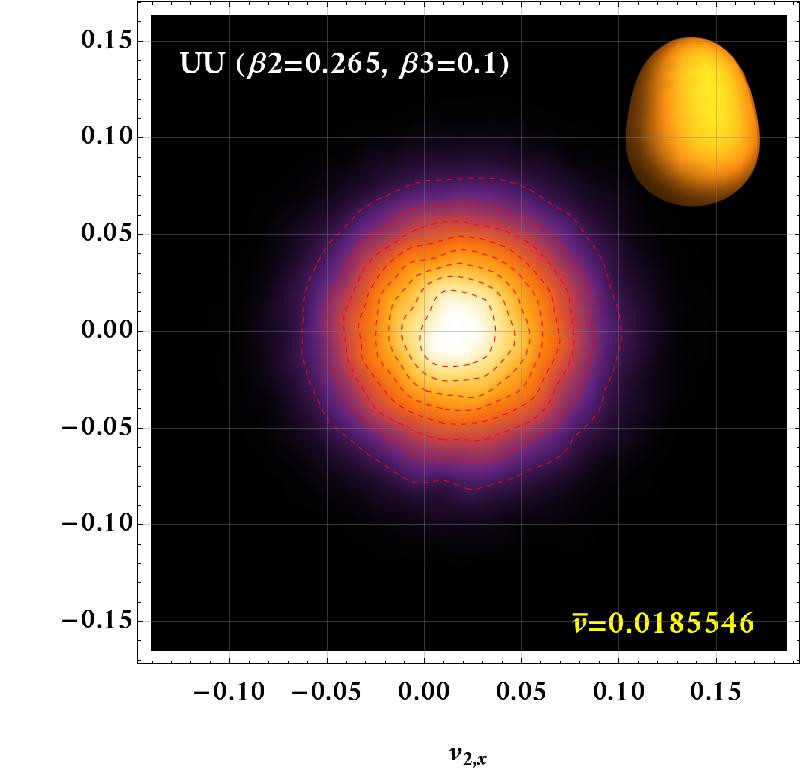}
	\end{tabular}		
	\caption{(Color online) These plots present $v_{2,x}=\alpha_2 \varepsilon_{2,x}$ vs. $v_{2,y}=\alpha_2 \varepsilon_{2,y}$ obtained from T$_{\text{R}}$ENTO using the same parametrization for different collisions.} 
	\label{fig9}
\end{figure}

\begin{figure}[h!]
	\begin{tabular}{c}
		\includegraphics[scale=0.4]{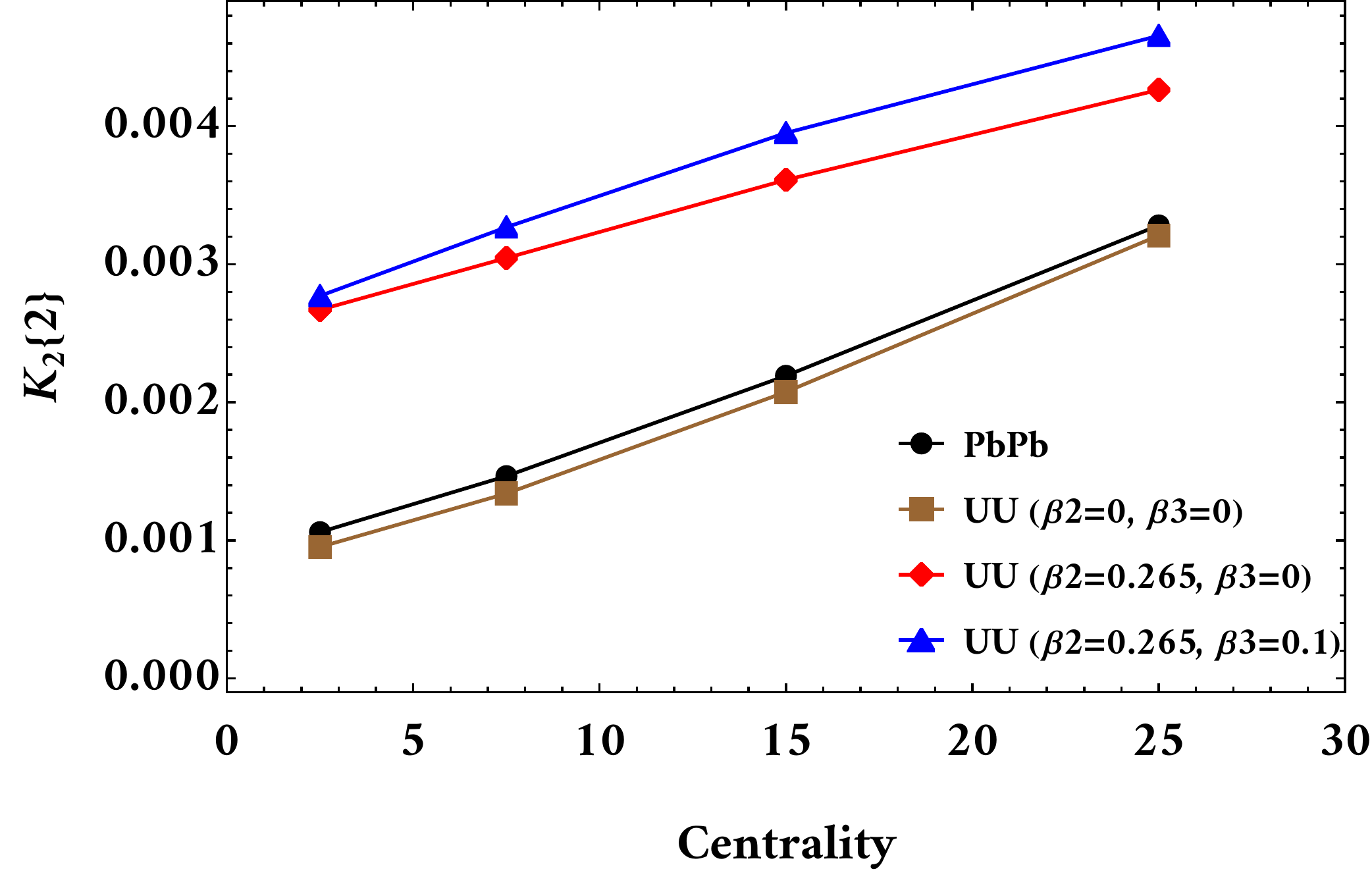}\\
		\includegraphics[scale=0.4]{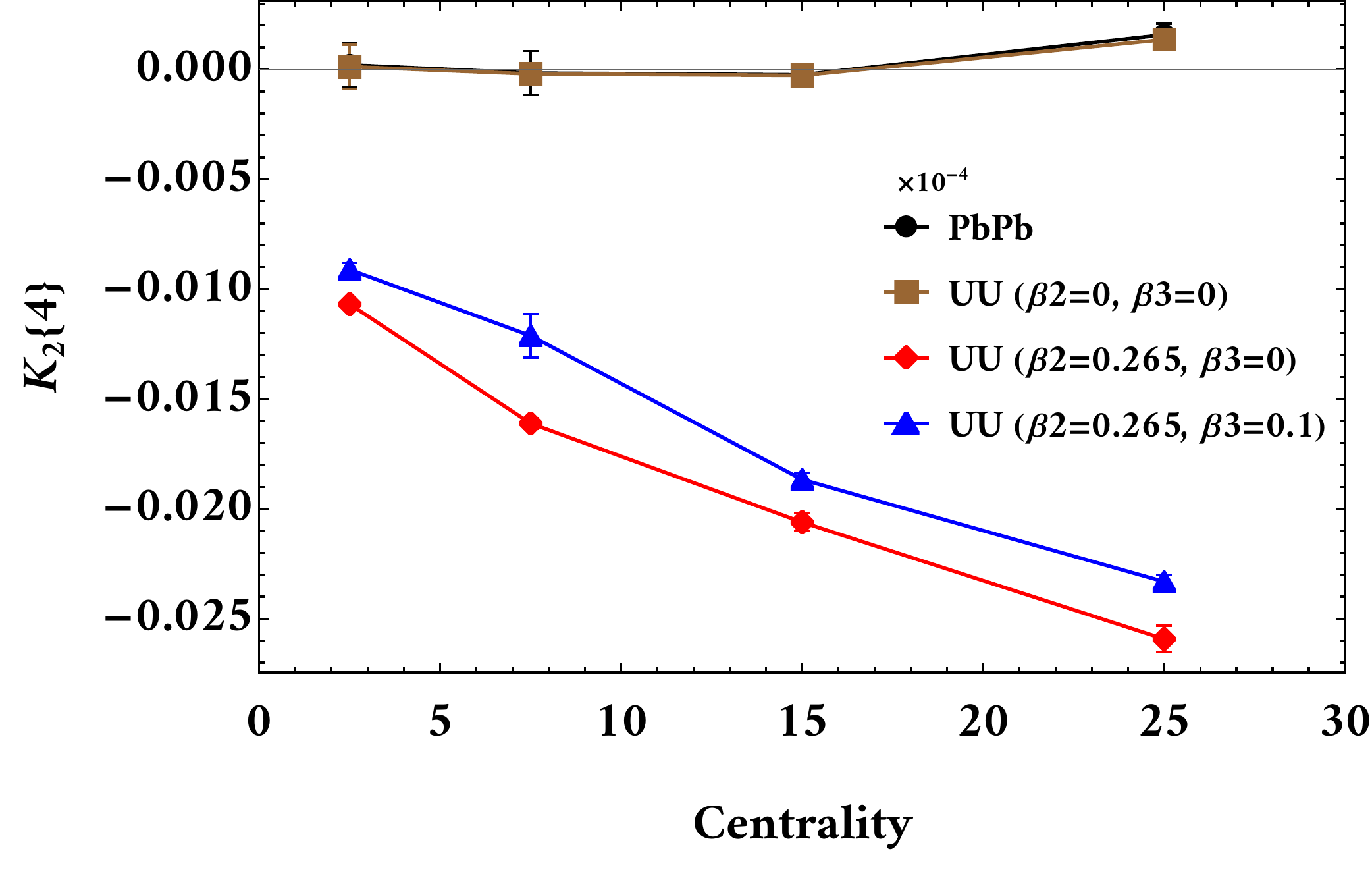}\\
		\includegraphics[scale=0.4]{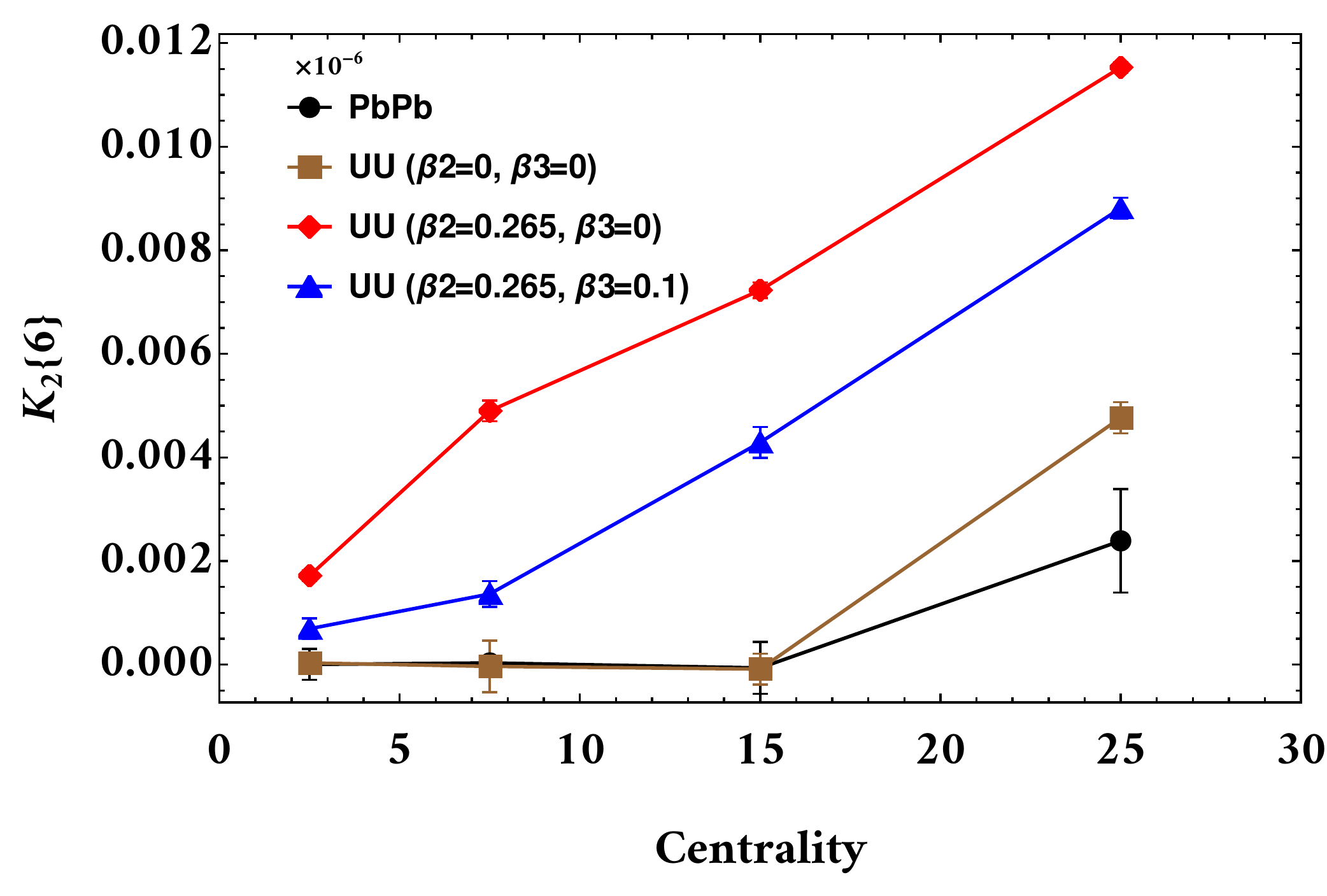}
	\end{tabular}		
	\caption{(Color online) This is similar to Fig.~\ref{fig1} for the shifted cumulants $K_n\{2k\}$. The results show that the relationship between different orders of shifted cumulants is $K_2\{2\}>K_2\{4\}>K_2\{6\}$ the same as $2k$-particle correlation functions.} 
	\label{fig2}
\end{figure}

In Fig.~\ref{fig9}, we present the 2-dimensional distributions of elliptic flow for different spherical and deformed nuclei collisions at $0-5\%$ centrality. The values of average ellipticity $\bar{v}_2$ are shown in each panel as well. Fig.~\ref{fig9} suggests that the flow distributions of different symmetric and asymmetric nuclei have the same behavior at most central collisions. Furthermore, it is known that the experimental data \cite{ATLAS:2019peb} favor Bessel-Gaussian (BG) distribution to explain elliptic flow distribution in spherical nuclei,
\begin{align*}
p(v_n)=\frac{2v_n}{c_n\{2\}}e^{-\frac{v_n^2+\bar{v}_n^2}{c_n\{2\}}}I_0\left(\frac{2v_n \bar{v}_n}{c_n\{2\}}\right),
\end{align*}
due to $c_n\{4\}\approx c_n\{6\}\approx\cdots\approx0$ at most central collisions. One may wonder if the BG can be a suitable choice for flow distribution of \emph{deformed ion} at low central collisions. The data of STAR \cite{STAR:2015mki} shows there is a noticeable difference between measured the $v_2$ of deformed and spherical nuclei collisions.   
This compels us to challenge our assumptions. As mentioned in Refs.\cite{hadi} and \cite{Mehrabpour:2020wlu}, we have to consider a shift in the x-direction, $Z=(v_{n,x}-\bar{v}_n)+iv_{n,y}$, where $\bar{v}_n=\la v_{n,x}\ra\neq0$ for even harmonics as depicted in Fig.~\ref{fig9}. However, this is not the case for odd harmonics where we have $\bar{v}_{2n+1}=0$. Imposing this change and following the previous steps, the relation between the generating functions of moments and cumulants of a complex variable is given by:
\begin{equation}\label{qq4}
\log \langle e^{t^*Z}\rangle =\sum_{k}\frac{(tt^*)^k}{(k!)}K_n\{2k\}.
\end{equation}
Therefore, one finds the desired order of the real cumulants $K_n\{2k\}$ \cite{Mehrabpour:2020wlu} as follows\footnote{It should be mentioned that they are derived by differentiating both sides of Eq.(\ref{qq4}) at $t_x=0$ and $t_y=0$. Also, we mention that $K_n\{2k\}$ are 2 dimensional cumulants.}:
\begin{equation}\label{qq5}
\begin{aligned}
K_n\{2\}&=\la ZZ^*\ra,\\
K_n\{4\}&=\la (ZZ^*)^2\ra-2\la ZZ^*\ra^2,\\
K_n\{6\}&=\la (ZZ^*)^3\ra+12\la ZZ^*\ra^3-9\la ZZ^*\ra\la (ZZ^*)^2\ra.
\end{aligned}
\end{equation}
Note that cumulants $K_n\{2k\}$ contain $c_n\{2k\}$ as well as moments such as $\bar{v}_n^i \la v_{n,x}^j v_{n,y}^l \ra$,  where $i+j+l=2k$ \footnote{By fixing the reaction plane in the experiment and computing an event-by-event $\mathbf{V}_n=v_ne^{in\Psi_n}=v_{n,x}+iv_{n,y}$ from the $\overrightarrow{\rm q}=(q_x,q_y)$ vectors, these cumulants can be found experimentally (see Ref.\cite{Bilandzic:2010jr} for more details). $v_n$ is the amplitude of anisotropic flow in the $n$-th harmonic, and $\Psi_n$ is the corresponding symmetry plane.}.
We call the above correlation functions, the shifted cumulants due to non-vanishing $\bar{v}_n$. 
In Fig.~\ref{fig2}, we plot Eq.~\ref{qq5} for $n=2$ for both spherical and deformed collisions. As depicted in the top plot ($K_2\{2\}$) of this figure, we observe a difference between spherical-spherical (SS) and deformed-deformed (DD) collisions. In addition, the effect of octupole deformation with non-zero $\beta_3$ in UU can be seen in this plot as well. For the particular cumulant $K_{2}\{2\}$, effect of $\beta_3$ manifests itself in mid-central collisions more noticeably. It turns out that the centrality dependence of cumulants for PbPb and spherical UU is very similar. Concerning the next order cumulants, i.e., $K_{2}\{4\}$, there is a considerable difference between the magnitude of the aforementioned quantity for deformed and spherical nuclei. We have $K_2\{4\}\approx 0$ for SS collisions as expected. The difference between deformed uranium collisions with and without $\beta_3$ manifest itself in $K_2\{4\}$ and $K_2\{6\}$. It can be seen that the effect of $\beta_3$ is decreasing in $K_{2}\{6\}$, whereas increasing in $K_{2}\{4\}$ and $K_{2}\{2\}$. This difference appears more clearly in $K_2\{6\}$ compared to the $K_2\{4\}$. As demonstrated in Figs.~\ref{fig1} and \ref{fig2}, the splitting between different values of quadrupole $\beta_2$ and octupole $\beta_3$ can be obtained from the shifted cumulants $K_n\{2k\}$ in contrast with $2k$-particle correlations $c_n\{2k\}$. This splitting appears stronger in the higher order of $K_n\{2k\}$. In other words, the results show that if we want to study the effect of deformation on flow anisotropies, it would be helpful to investigate the shifted cumulants $K_n\{2k\}$. Also, the difference between different spherical nuclei at mid-central centrality can be extracted in $K_2\{6\}$. To see this difference, one needs to include higher-order terms of cumulants in probability distributions, which we leave to future work.

As we observed the correction $\bar{v}_n$ in our assumptions led us to extract some fascinating properties. Now, we study the effect of this modification on flow distributions. 
As discussed in Sec.~\ref{sec2}, one can study a stochastic variable using the sGC method and its characteristic function, which gives the modification to the Gaussian distribution. Thus, using Eq.~\ref{qq4}, we can investigate shifted flow harmonics $Z=v_{n,x}-\bar{v}_n+iv_{n,y}$. To have an approach accessible in experiments, we should obtain a distribution for the magnitudes of flow harmonics. To do this, we find the radial distribution $P_r(v_n)$ by writing the distribution given in Cartesian coordinate in the polar coordinate and then integrating the two-dimensional distribution $P(v_{n,x},v_{n,y})$ over the $\Psi_n$ (see Ref.\cite{Mehrabpour:2020wlu}) as follows:
\begin{equation}\label{qq6}
\begin{split}
P_r(v_n)
&=\frac{d}{dv_n}\int P(v_x,v_y)\;dv_x\;dv_y\\
&=\frac{d}{dv_n}\int v_n\; P(v_n,\Psi_n)\;dv_n\;d\Psi_n,
\end{split} 
\end{equation}
where 
\begin{equation}\label{qq7} 
P(v_n,\Psi_n)\approx\left[1+\sum_{k=2} \frac{R_n\{2k\}\mathcal{D}^{k}_{v_n,\Psi_n}}{4^k(k!)^2}\right] \mathcal{G}(v_n,\Psi_n).
\end{equation}
Here, $\sqrt{2\pi}\sigma \mathcal{G}(z,\phi)$ is a 2D Gaussian distribution with mean $\bar{v}_n$ and standard deviation $\sqrt{R_n\{2\}/2}$. Moreover, $\mathcal{D}=\partial_{v_n}^2+(1/v_n)\partial_{v_n}+(1/v_n^2)\partial_{\Psi_n}^2$.

\begin{figure}[t!]
	\begin{tabular}{c}
		\includegraphics[scale=0.42]{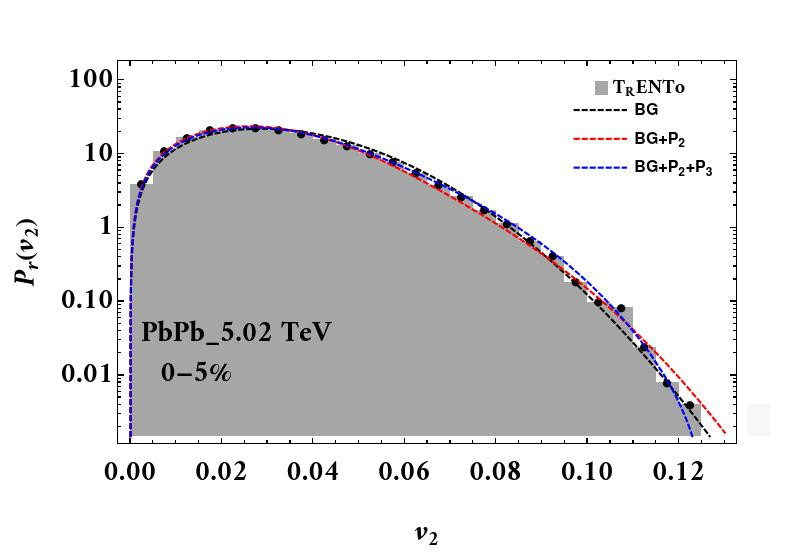}\\
		\includegraphics[scale=0.42]{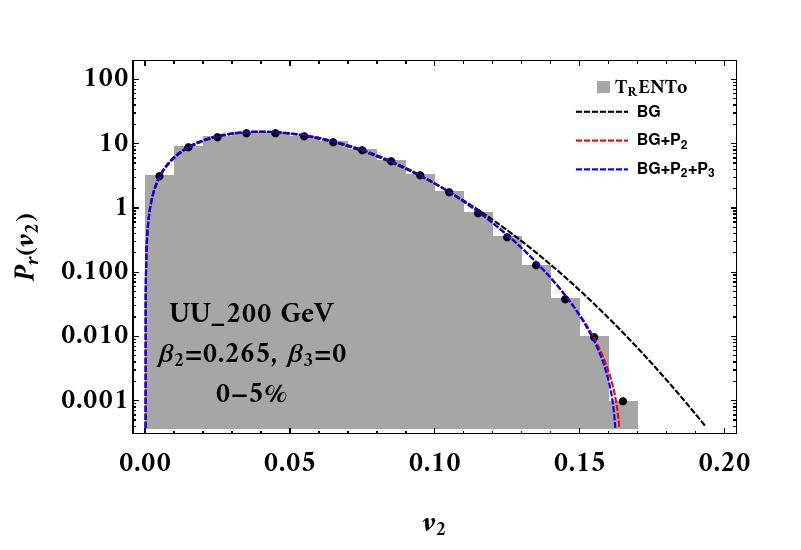}
	\end{tabular}		
	\caption{(Color online) A comparison of the obtained elliptic distribution with $BG(v_2)$ and different corrections $P_2$ and $P_3$ is presented by dashed black, red, and blue lines, respectively. The top panel displays the data of PbPb at $0-5\%$, while the bottom panel shows the results of UU collisions in the same centrality class.} 
	\label{fig3}
\end{figure}

\begin{figure}[t!]
	\begin{tabular}{c}
		\includegraphics[scale=0.45]{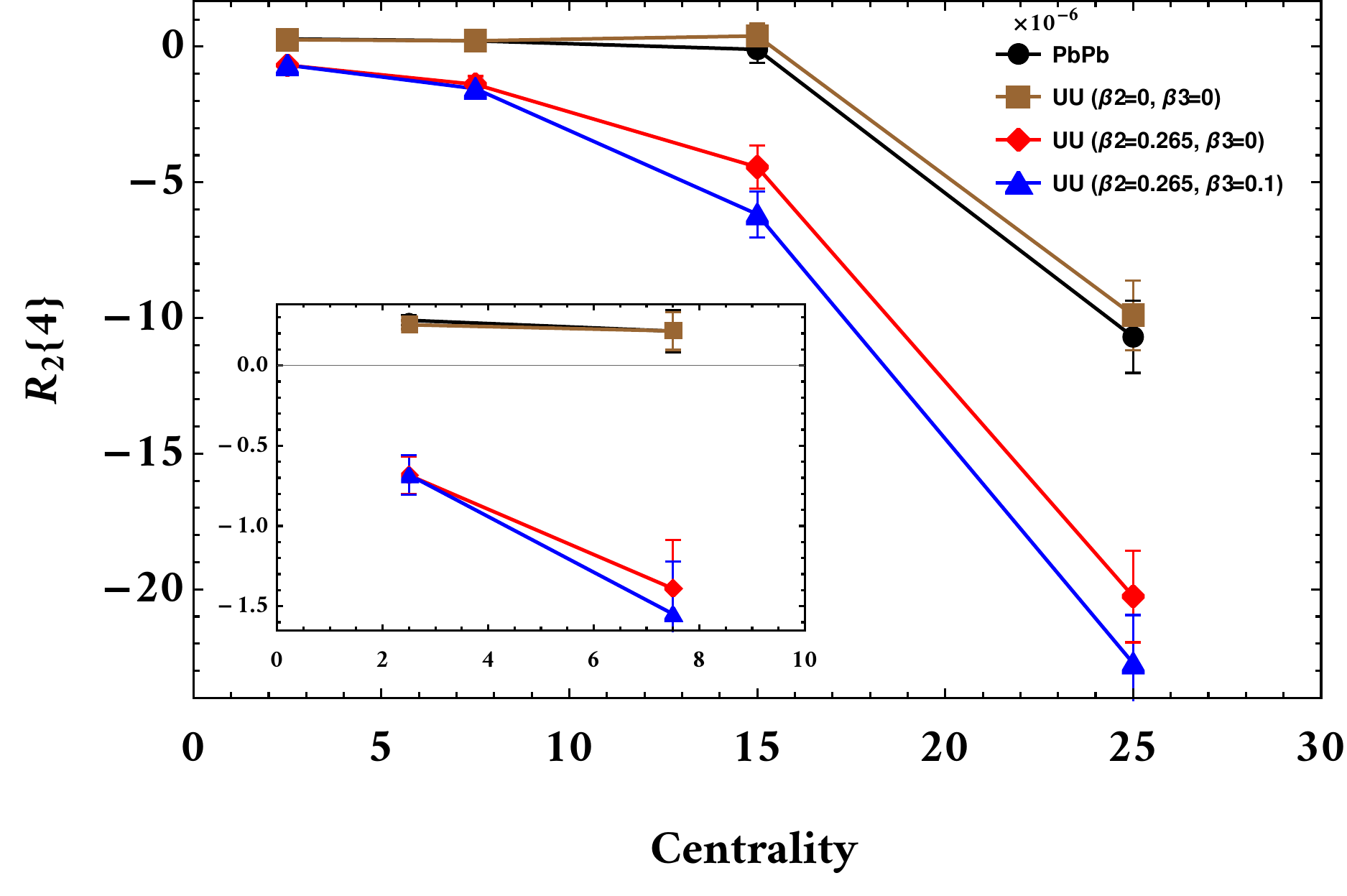}
	\end{tabular}		
	\caption{(Color online) Comparing the cumulants $R_2\{4\}$ as a function of centrality for different spherical and deformed nuclei collisions. The mini panel presents the values $R_2\{4\}$ where we expect maximum deformity.} 
	\label{fig4}
\end{figure}

Bearing in mind the importance of the shift parameter, introduced in Eq.~\ref{qq5}, we have to use the radial shifted cumulants $R_n\{2k\}$. This means that the cumulants in Eq.~\ref{qq4} are not applicable anymore. Therefore, we employ the main definition of moments,
\begin{equation}\label{q7}
\la v_n^{2k}\ra= \int_{0}^{\infty} v_n^{2k} P_r(v_n) \;dv_n,
\end{equation}
to obtain $R_n\{2k\}$ (see Ref.\cite{hadi}).
In order to obtain an expression for $R_{n}\{2k\}$, we truncate Eq.~\ref{qq7} at desired order of $k$. For example, if we keep only the first term in Eq.~\ref{qq7}, $P_r(v_n)$ would be a BG distribution:
\begin{equation}\label{qq8}
P_r(v_n)\equiv BG(v_n)=\mathcal{G}(v_n;\bar{v}_n)I_0\left(\frac{2v_n\bar{v}_n}{R_n\{2\}}\right),
\end{equation}       
where $\mathcal{G}(v_n;\bar{v}_n)=(2v_n/R_n\{2\})\exp\left[-\frac{v_n^2+\bar{v}_n^2}{R_n\{2\}}\right]$ is 1D Gaussian distribution with non-zero central moment, and $I_j(z)$ is the modified Bessel function of the first kind. In this case, we just have $R_n\{2\}$:
\begin{align*}
\la v_n^2\ra&= \int v_n^2 P_r(v_n)\;dv_n=\int v_n^2 BG(v_n)\;dz=R_n\{2\}+\bar{v}_n^2.
\end{align*}
Now, we obtain the form of the first shifted cumulants using the equality above as $R_n\{2\}=\la v_n^2\ra-\bar{v}_n^2=c_n\{2\}-\bar{v}_n^2$. However, we can keep higher order terms in this expansion as well to arrive at a distribution that includes corrections to Bessel-Gaussianity in Eq.~\ref{qq8} as $P_r(v_n)=BG(v_n)+P_k(v_n)$. Here, $P_k(v_n)$ includes corrections from higher cumulants. We know from experiments \cite{Aad:2014vba} that BG distribution presents a suitable description of the flow harmonics of spherical nuclei from most to mid-central collisions. Now, one may wonder if we can use this expression for deformed nuclei in the same centrality classes. To answer this, we consider the correction to BG distribution which just includes the second and third radial shifted cumulants, $R_n\{4\}$ and $R_n\{6\}$, as follows:      
\begin{equation}\label{qqq8}
\begin{split}
P_r(v_n)&=BG(v_n)+P_2(v_n)+P_3(v_n)
\\&=\mathcal{G}(v_n;\bar{v}_n)I_0\left(\frac{2v_n\bar{v}_n}{R_n\{2\}}\right) \\&+\frac{1}{2}\gamma_4\mathcal{G}(v_n;\bar{v}_n)\sum_{j=0}^{2}\alpha_{2,j} I_j\left(2v_n\bar{v}_n/R_n\{2\}\right)
\\&+\frac{1}{6}\gamma_6\mathcal{G}(v_n;\bar{v}_n)\sum_{j=0}^{3}\alpha_{3,j} I_j\left(2v_n\bar{v}_n/R_n\{2\}\right).
\end{split}	
\end{equation}     
The coefficients $\gamma_{2k}$ and $\alpha_j$ in the correction terms of Eq.~\ref{qqq8} are:
\begin{align*}
\gamma_4&=R_n\{4\}/R_n\{2\}^2,\;\gamma_6=R_n\{6\}/R_n\{2\}^3,\\
\alpha_{2,0}&=\mathit{L}_2 \left(\frac{v_n^2+\bar{v}_n^2}{R_n\{2\}}\right)+\frac{v_n^2\bar{v}_n^2}{R_n\{2\}^2},\\
\alpha_{3,0}&=\mathit{L}_3 \left(\frac{v_n^2+\bar{v}_n^2}{R_n\{2\}}\right)+\frac{3v_n^2\bar{v}_n^2}{R_n\{2\}^2}\mathit{L}_1\left(\frac{v_n^2+\bar{v}_n^2}{3
	R_n\{2\}}\right),\\
\alpha_{2,1}&=\frac{4v_n\bar{v}_n}{R_n\{2\}}\mathit{L}_1 \left(\frac{v_n^2+\bar{v}_n^2}{2R_n\{2\}}\right),\\
\alpha_{3,1}&=\frac{6v_n\bar{v}_n}{R_n\{2\}}\mathit{L}_2 \left(\frac{v_n^2+\bar{v}_n^2}{2R_n\{2\}}\right)+\frac{v_n^4+6v_n^2\bar{v}_n^2+\bar{v}_n^4}{8R_n\{2\}^2},\\
\alpha_{2,2}&=\frac{v_n^2\bar{v}_n^2}{R_n\{2\}^2},\;
\alpha_{3,2}=\frac{3v_n^2\bar{v}_n^2}{R_n\{2\}^2}\mathit{L}_1\left(\frac{v_n^2+\bar{v}_n^2}{3
	R_n\{2\}}\right),\\
\alpha_{3,3}&=\frac{v_n^3\bar{v}_n^3}{3R_n\{2\}^3}.
\end{align*}  
Here, $\mathit{L}_i(z)$ are the Laguerre polynomials. Let us emphasize that we kept up to the third orders $k=3$ since higher order cumulants are small \footnote{To examine this claim, we included them and confirmed they are small and negligible.}. Plugging, Eq.~\ref{qqq8} into Eq.~\ref{q7}, the cumulants $R_{n}\{2k\}$ with $k=1,2,3$ are given by: 
\begin{equation}\label{qR}
\begin{split}
R_n\{2\}&=\la v_n^2\ra-\bar{v}_n^2=c_n\{2\}-\bar{v}_n^2,\\
R_n\{4\}&=\la v_n^4\ra-2\la v_n^2\ra^2+\bar{v}_n^4=c_n\{4\}+\bar{v}_n^4.\\
R_n\{6\}&=\la v_n^6\ra-9\la v_n^4\ra\la v_n^2\ra+12\la v_n^2\ra^3-4\bar{v}_n^6=c_n\{6\}-4\bar{v}_n^6.
\end{split}
\end{equation} 
We see that $R_n\{2\}=K_n\{2\}$ as expected. So, if one obtains $R_n\{2\}$ for different collisions, the results in the top panel of Fig.~\ref{fig2} are reproduced. To investigate $P_r(v_n)$ for both SS and DD collisions, we focus on $0-5\%$ centrality where we expect to have the most deformity of nuclei \cite{STAR:2015mki}. Fig.~\ref{fig3} shows a comparison of the distribution of PbPb with UU. As depicted in this figure, the leading order truncation of Eq.~\ref{qq8} is a reliable estimation for PbPb data in most central collisions. In contrast to PbPb, the distribution of UU indicates a trace of non-Bessel-Gaussianity. This comes from the term involving $R_2\{4\}$ comparable to the leading term in deformed UU collisions. In this context, Fig.~\ref{fig4} shows that there is a noticeable difference between the values of $R_2\{4\}$ for spherical and deformed collisions. Furthermore, once $\beta_3$ is turned on, we can observe an increase in the magnitude of $R_2\{4\}$ opposite to $c_2\{4\}$ and $K_2\{4\}$. However, there is a splitting between different deformed nuclei as well.
\begin{figure}[t!]
	\begin{tabular}{c}
		\includegraphics[scale=0.42]{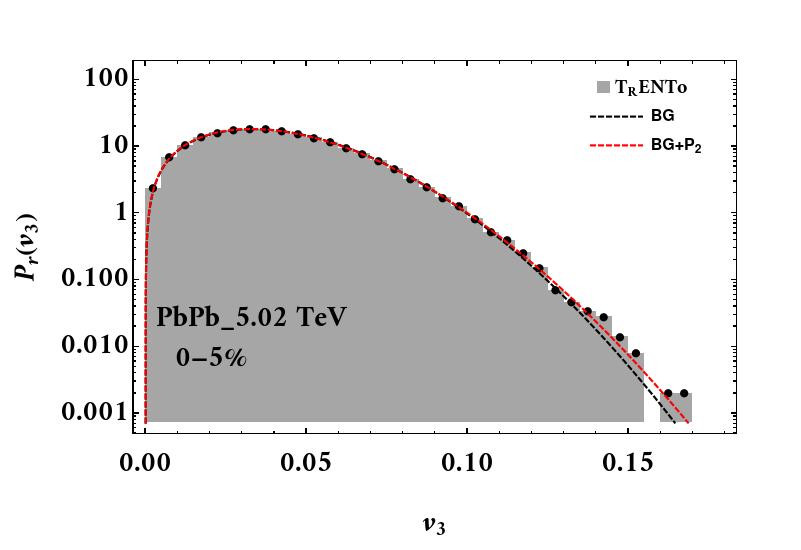}\\
		\includegraphics[scale=0.42]{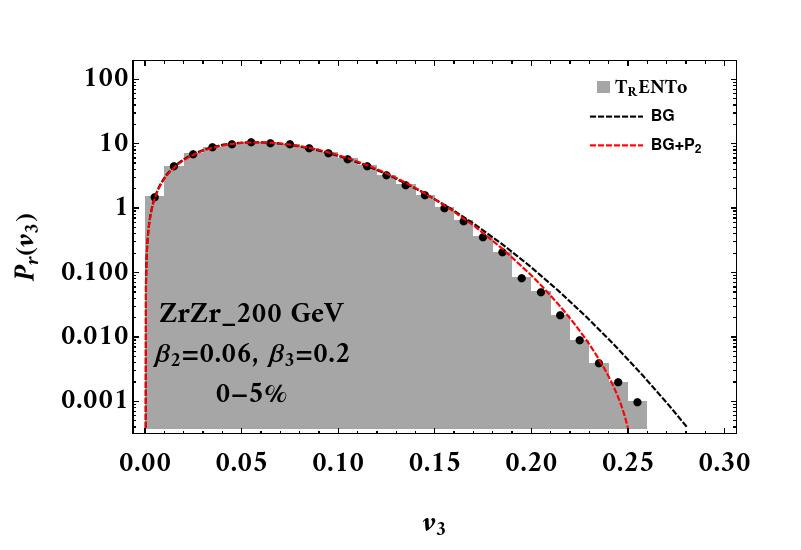}
	\end{tabular}		
	\caption{(Color online) Similar to Fig.~\ref{fig3} but for the third harmonic $v_3$ distribution of PbPb (top) and ZrZr (bottom) collisions.} 
	\label{fig5}
\end{figure}
If we want to investigate the deformation effect on $v_3$ or the octupole structure of nuclei, it seems that increasing $\beta_3$ would lead to a correction to Bessel-Gaussianity as well. In Fig.~\ref{fig5}, we show the distribution of $v_3$ both for SS (PbPb) and DD (ZrZr) collisions in $0-5\%$ centrality. The results imply that large values of $\beta_3$ play a significant role in $v_3$ distribution. This effect appears as a non-Bessel-Gaussian distribution, while the Bessel-Gaussian approximation works well for spherical nuclei.        
\begin{figure}[t!]
	\begin{tabular}{c}
		\includegraphics[scale=0.445]{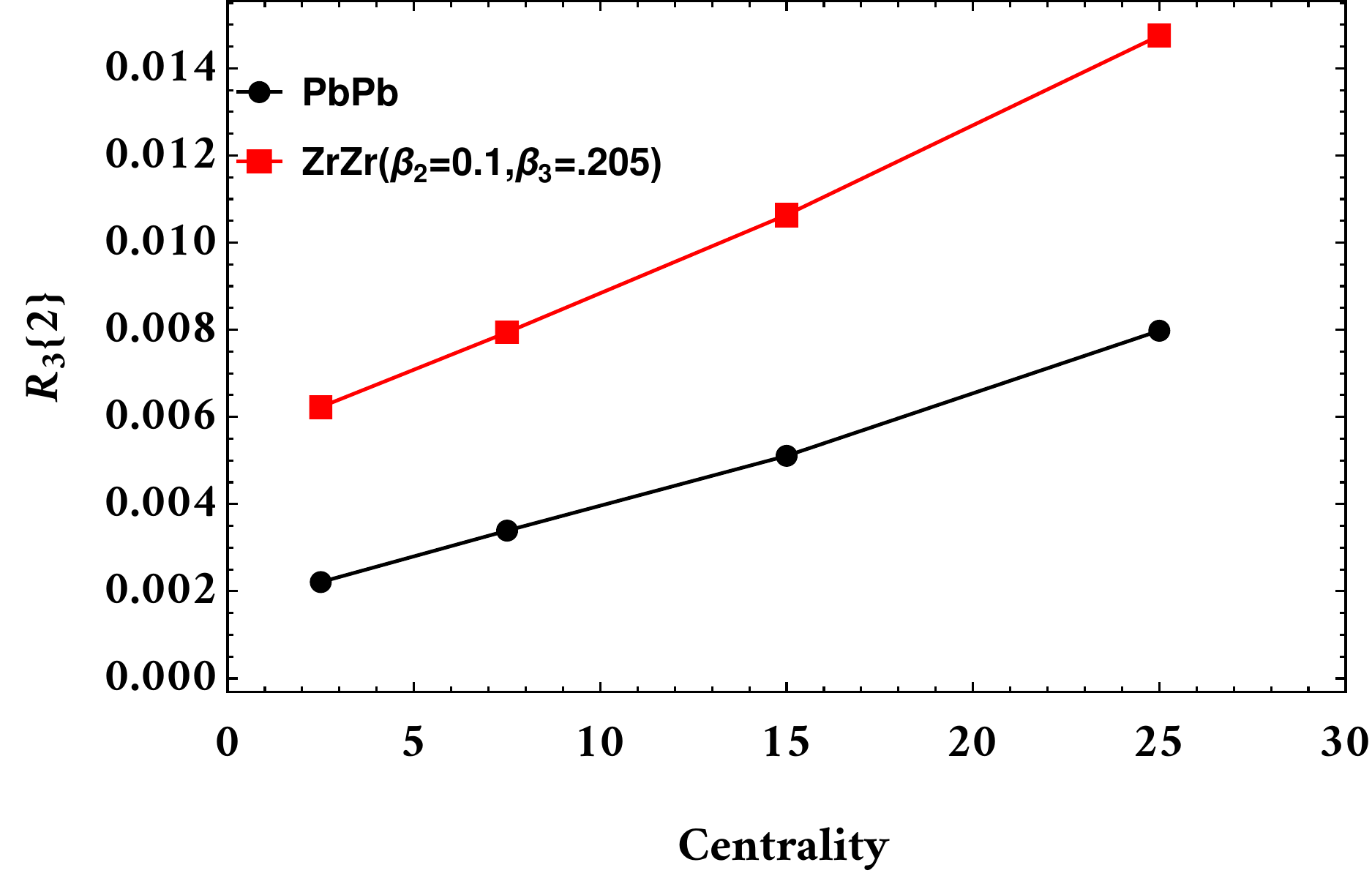}\\
		\includegraphics[scale=0.45]{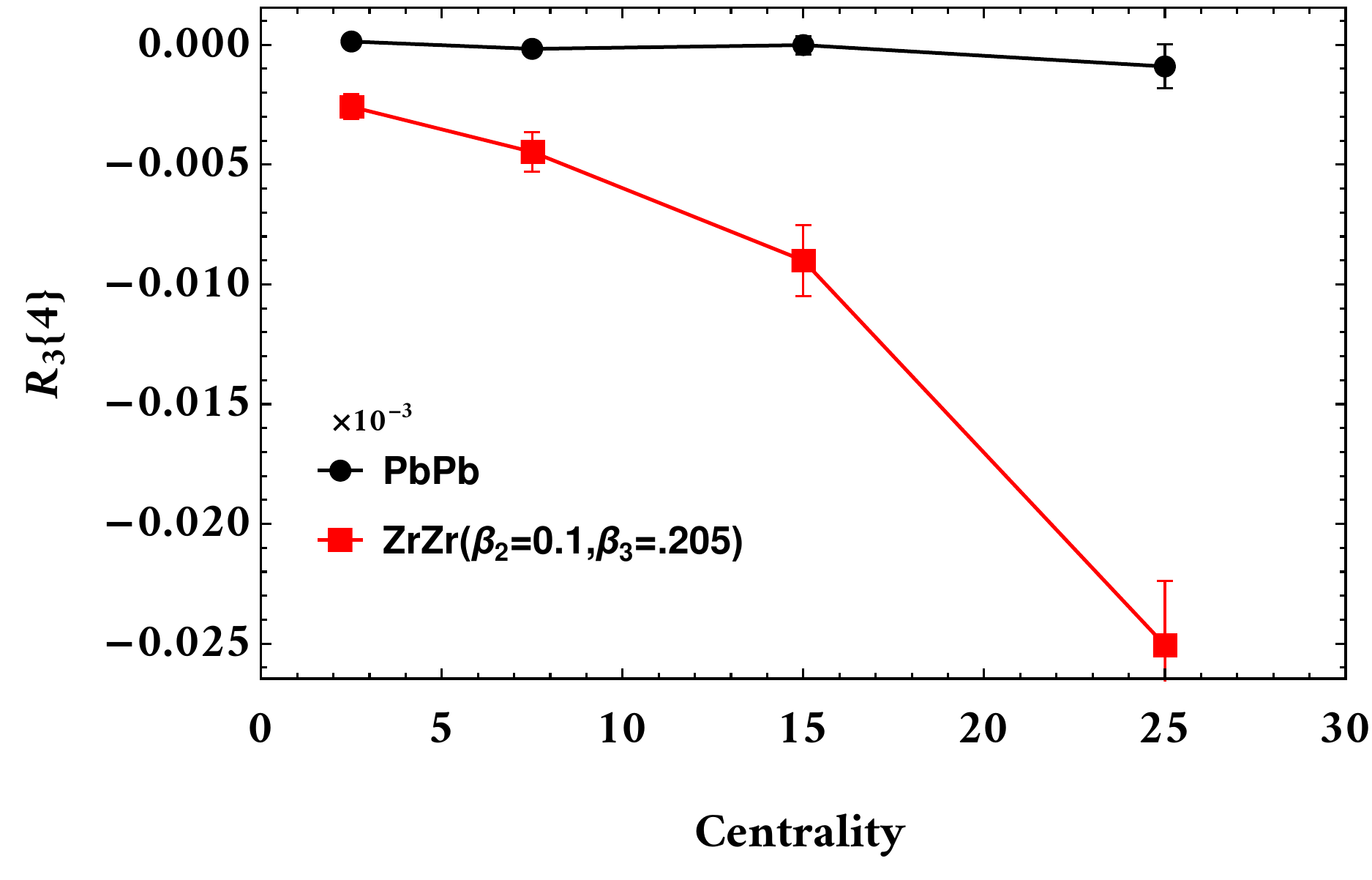}
	\end{tabular}		
	\caption{(Color online) The results show the values of $R_3\{2\}$ and $R_3\{4\}$, which obtained from PbPb data and ZrZr data.} 
	\label{fig6}
\end{figure}
To study the correction part of $v_3$ distribution, we should investigate the coefficient $\gamma_4=R_3\{4\}/R_3\{2\}^2$. Fig.~\ref{fig6} shows a comparison of $R_3\{2\}$ and $R_3\{4\}$ for PbPb with ZrZr collisions. Note that due to $\bar{v}_3=0$, one can find $R_3\{2k\}=c_3\{2k\}$. As illustrated in Fig.~\ref{fig6}, we find that the cumulants of ZrZr have a larger magnitude than spherical ones. This leads to a non-negligible difference in $\gamma_4$, and thus, the correction terms are crucial in ZrZr as shown in Fig.~\ref{fig5}. The deformation of nuclei manifests itself in the shape of colliding nuclei. On the other hand, the change in the overlapping area can have effects on the cumulants and distributions of flow harmonics. In this regard, we believe that the standard Gram-Charlier series would be ideal tool as a probe of nuclear structure in distribution analysis.      

\section{relation between observables}\label{sec5}
In Sec.~\ref{sec3}, we presented the comparison of SS and DD distributions as a probe of nuclear structure. We showed that the effect of deformation appears for the second and third harmonics. It is useful to have an estimate of the observable DD collisions. Thus, we want to estimate the observables by fitting a known SS distribution, e.g., PbPb, to the deformed nuclei data like UU. To do that, we need to modify the distribution of the spherical one. Since the correction included at $k=3$ is negligible, we modify $P_r(v_n)$ in Eq.~\ref{qqq8} by considering the truncation at $k=2$:  
\begin{equation}\label{qq9}
\begin{split} 
P_r^{M}&(v_n)=\mathcal{G}(v_n';\bar{v}_{est})I_0\left(2v_n'\bar{v}_{est}/R_n\{2\}_{est}\right)\\
&+\frac{1}{2}\gamma_4^{est}\mathcal{G}(v_n';\bar{v}_{est})\sum_{j=0}^{2}\alpha_{2,j}^{est}(v_n') I_j\left(2v_n'\bar{v}_{est}/R_n\{2\}_{est}\right),
\end{split} 	 
\end{equation}
Inspired by Refs.\cite{Jia:2021tzt} and \cite{Jia:2021qyu}, the estimated parameters in the above are defined by: 
\begin{equation}\label{qq10}
\begin{split}
&v_n'=v_{n,0}+\sum_{m=2}p_m\beta_m,\quad 
\bar{v}_{est}=\bar{v}_{0}+\sum_{m=2}\delta_{1,m} \beta_m,\\
&R_n\{2\}_{est}=R_n\{2\}_{0}+\left(\sum_{m=2}\delta_{2,m} \beta_m\right)^2,\\
&R_n\{4\}_{est}=R_n\{4\}_{0}+\left(\sum_{m=2}\delta_{3,m} \beta_m\right)^4,
\end{split}
\end{equation}
\begin{figure}[t!]
	\begin{tabular}{c}
		\includegraphics[scale=0.42]{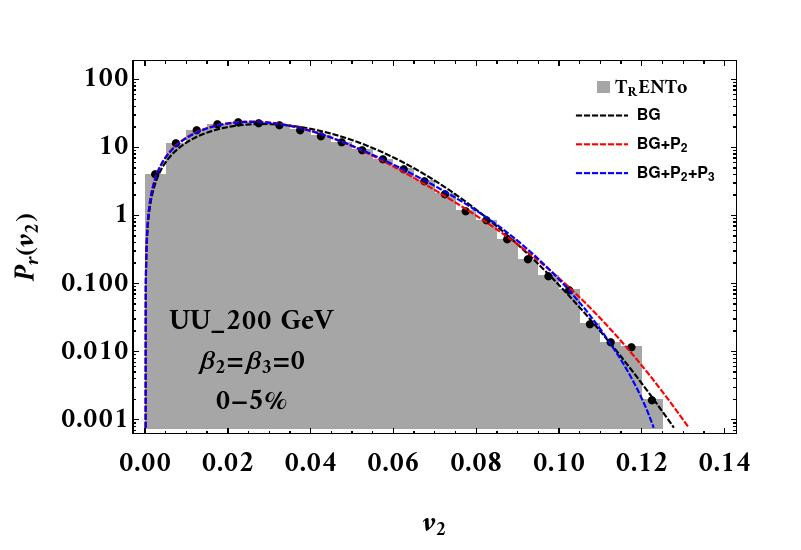}\\
		\includegraphics[scale=0.42]{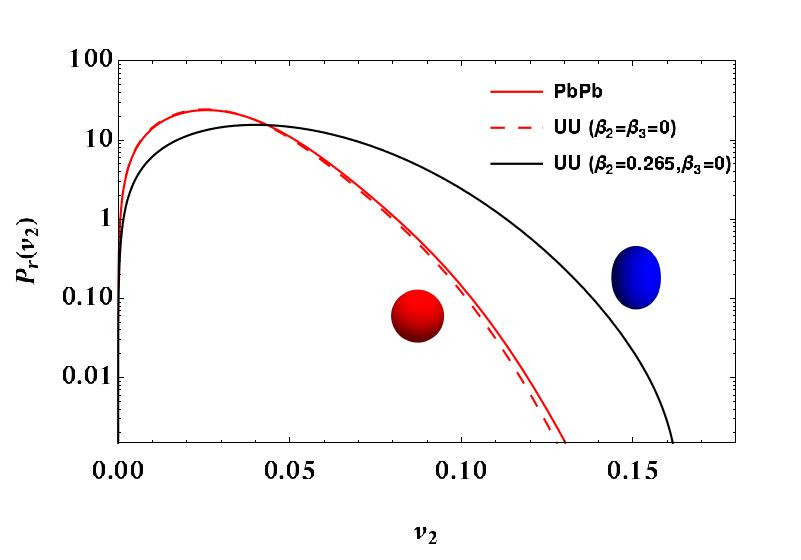}\\
		\includegraphics[scale=0.42]{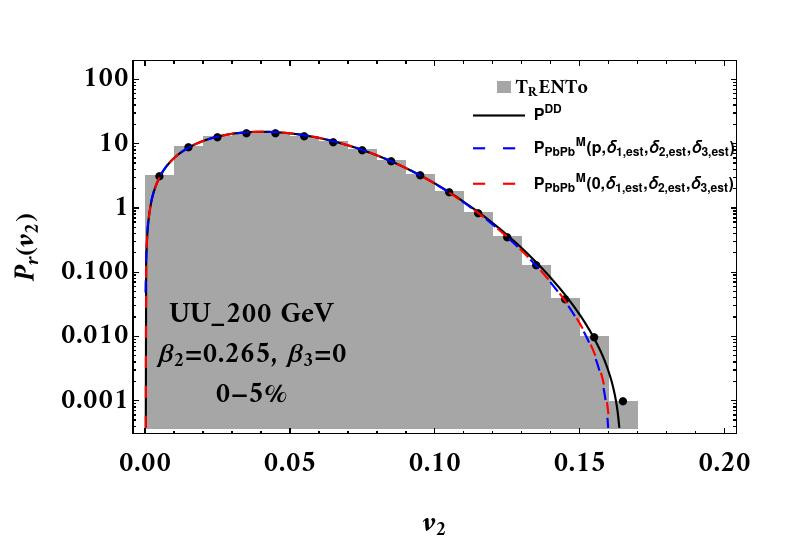}
	\end{tabular}		
	\caption{(Color online) In the top panel, the different corrections of spherical uranium in most-central collisions are compared; while in the middle panel, the distributions of spherical nuclei are compared to the flow distribution of deformed uranium collisions. The bottom panel shows different estimations of DD using the SS distribution.} 
	\label{fig7}
\end{figure}where the index $0$ in the above indicates spherical observables. 		
In fact, Eq.~\ref{qq10} is the simplest case to study the impact of deformation directly in terms of observables. However, this modification allows us to study the effect of deformation directly, in analogy to Ref.\cite{Jia:2021tzt}. Here, we show that having the cumulants obtained from PbPb data, one arrives at the UU observables using Eqs.~\ref{qq9} and \ref{qq10}. To do this, we show that the distribution of spherical PbPb and UU is equivalent. In the top panel of Fig.~\ref{fig7}, we plot the distribution of spherical uranium with the vanishing deformation parameter $\beta_{2}=\beta_{3}=0$. It is obvious that the BG distribution can explain the data for the spherical uranium accurately. In the middle panel of this plot, the comparison of PbPb and spherical uranium shows good agreement between them. This allows us to estimate the deformed UU observables using PbPb data. Since we want to study quadrupole deformation of nuclei, as a simple case study, we generate UU collisions by setting $\beta_2=0.265$ and $\beta_3=0$. This is because of removing the $\beta_3$ effect on the cumulants $R_2\{2k\}$. As illustrated in the middle plot of Fig.~\ref{fig7}, there is a noticeable difference between SS and DD distributions. It should be mentioned that truncation at $k=2$ was considered for both SS and DD.  
\begin{figure}[t!]
	\begin{tabular}{c}
		\includegraphics[scale=0.42]{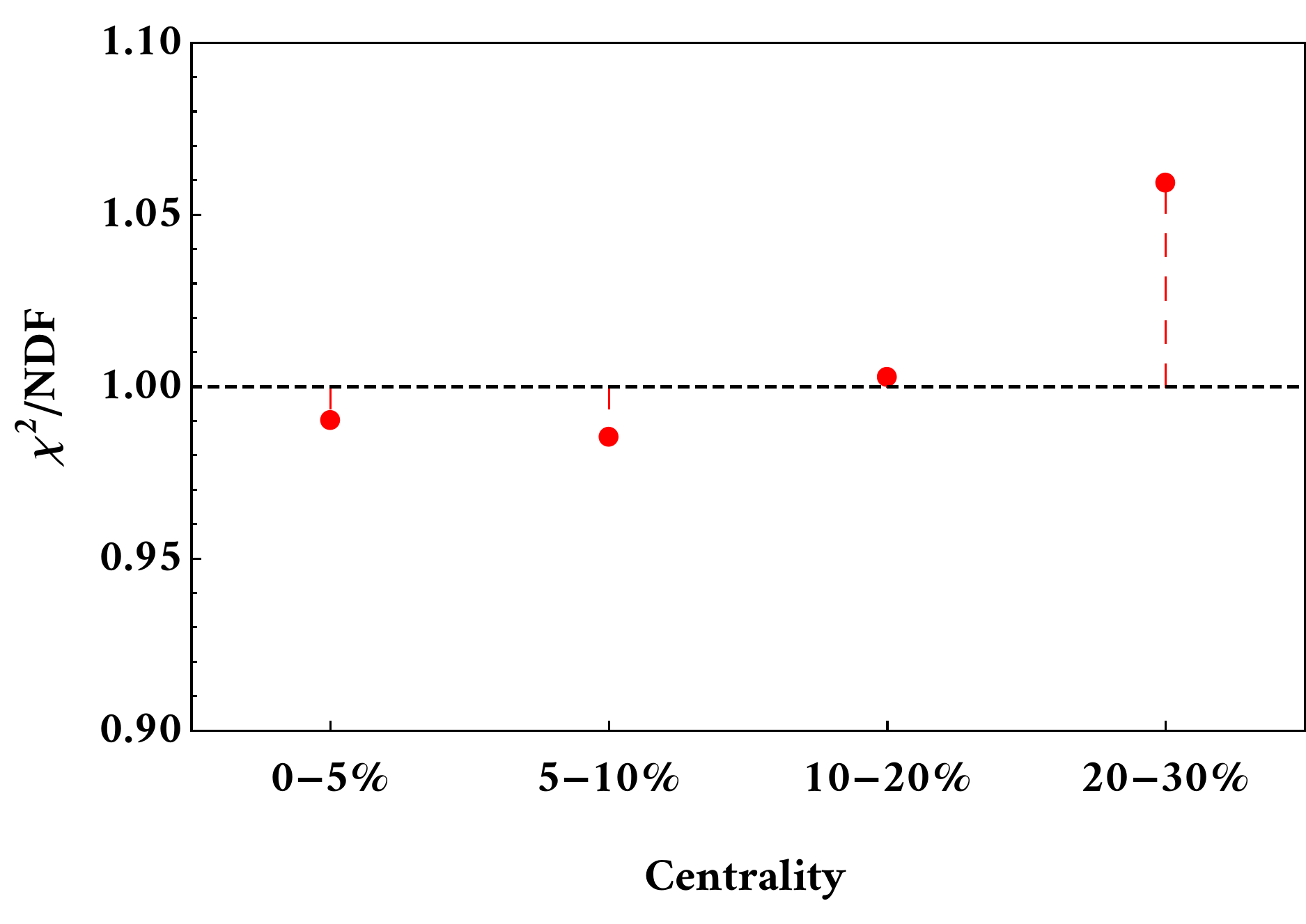}
	\end{tabular}		
	\caption{(Color online) $\chi^2/$NDF values of fitting the distribution $P^M_r(v_2)$ to simulation data as function of centrality.} 
	\label{fig8}
\end{figure}
Finding the coefficients $p$ and $\delta_i$ leads us to the distribution of deformed nuclei. Since we just considered non-zero $\beta_2$, we rename the coefficients in Eq.~\ref{qq10} as $p=p_2$, $\delta_{1,est}=\delta_{1,2}$, $\delta_{2,est}=\delta_{2,2}^2$, and $\delta_{3,est}=\delta_{3,2}^4$. As demonstrated in the bottom panel of Fig.~\ref{fig7}, we found different estimations of UU distribution as follows:
\begin{equation}\label{qq11}
\begin{split}
P^{DD}_{est}&=P^{M}(p,\delta_{1,est},\delta_{2,est},\delta_{3,est}),\\
P^{DD}_{est}&=P^{M}(0,\delta_{1,est},\delta_{2,est},\delta_{3,est}),
\end{split}	
\end{equation}
Results show that the estimated distributions are compatible with the $v_2$ distribution obtained from UU data qualitatively. In other words, the definitions in Eq.~\ref{qq10} worked. To find their consistency, one can investigate them in other centralities. The values of coefficients $\delta_{i,est}$ are presented in Table.~\ref{tab1} at different centrality classes. We can see that the effect of deformation would be different in each centrality due to various estimated values.
Also, $\chi^2/$NDF of fitting in each centrality are illustrated in Fig.~\ref{fig8}. Since these values are closer to 1, one can interpret that Eqs.~\ref{qq9} and \ref{qq10} present a good estimation of deformed UU observables. Of course, the value of $\chi^2/$NDF in mid-central collisions is growing. This means that that if we go to higher centralities we needs to other truncations in Eq.~\ref{qq7}, e.g. keep the terms included $R_n\{6\}$,$R_n\{8\}$ and so on to explain data without any fitting. Moreover, if one wants to study $2k$-particle correlation functions $c_n\{2k\}$ in this context, they should be written as a function included $\beta_n$:
\begin{equation}\label{qcn} 
\begin{split}
c_n\{2k\}_{est}&=c_n\{2k\}_0+\left(\sum_{m=2}\xi_{2k,m} \beta_m\right)^{2k}. 
\end{split}
\end{equation} 
As mentioned in Eq.~\ref{qR}, $R_n\{2k\}_{est}$ is a function of $c_n\{2k\}_{est}$ and $\bar{v}_{n,est}$. Plugging Eq.~\ref{qcn} in Eq.~\ref{qR} and separating the terms with $\beta_n$ from spherical terms, one can find the following relations:
\begin{equation}\label{qxi}
\begin{split}
R_n&\{2\}_{est}=R_n\{2\}_{0}-\left(\sum_{m=2}\delta_{1,m} \beta_m\right)^2\\&-2\left(\sum_{m=2}\delta_{2,m} \beta_m\right)\bar{v}_0+\left(\sum_{m=2}\xi_{2,m} \beta_m\right)^{2},\\
R_n&\{4\}_{est}=R_n\{4\}_{0}+\left(\sum_{m=2}\delta_{1,m} \beta_m\right)^4\\&+4\left(\sum_{m=2}\delta_{1,m} \beta_m\right)^3\bar{v}_0+6\left(\sum_{m=2}\delta_{1,m} \beta_m\right)^2\bar{v}_0^2\\&+4\left(\sum_{m=2}\delta_{1,m} \beta_m\right)\bar{v}_0^3+\left(\sum_{m=2}\xi_{4,m} \beta_m\right)^{4},
\end{split}	
\end{equation}
keeping in mind $R_n\{2\}_0=c_n\{2\}_0-\bar{v}_{n,0}^2$ and $R_n\{4\}_0=c_n\{4\}_0+\bar{v}_{n,0}^4$.
\begin{figure}[t!]
	\begin{tabular}{c}
		\includegraphics[scale=0.4]{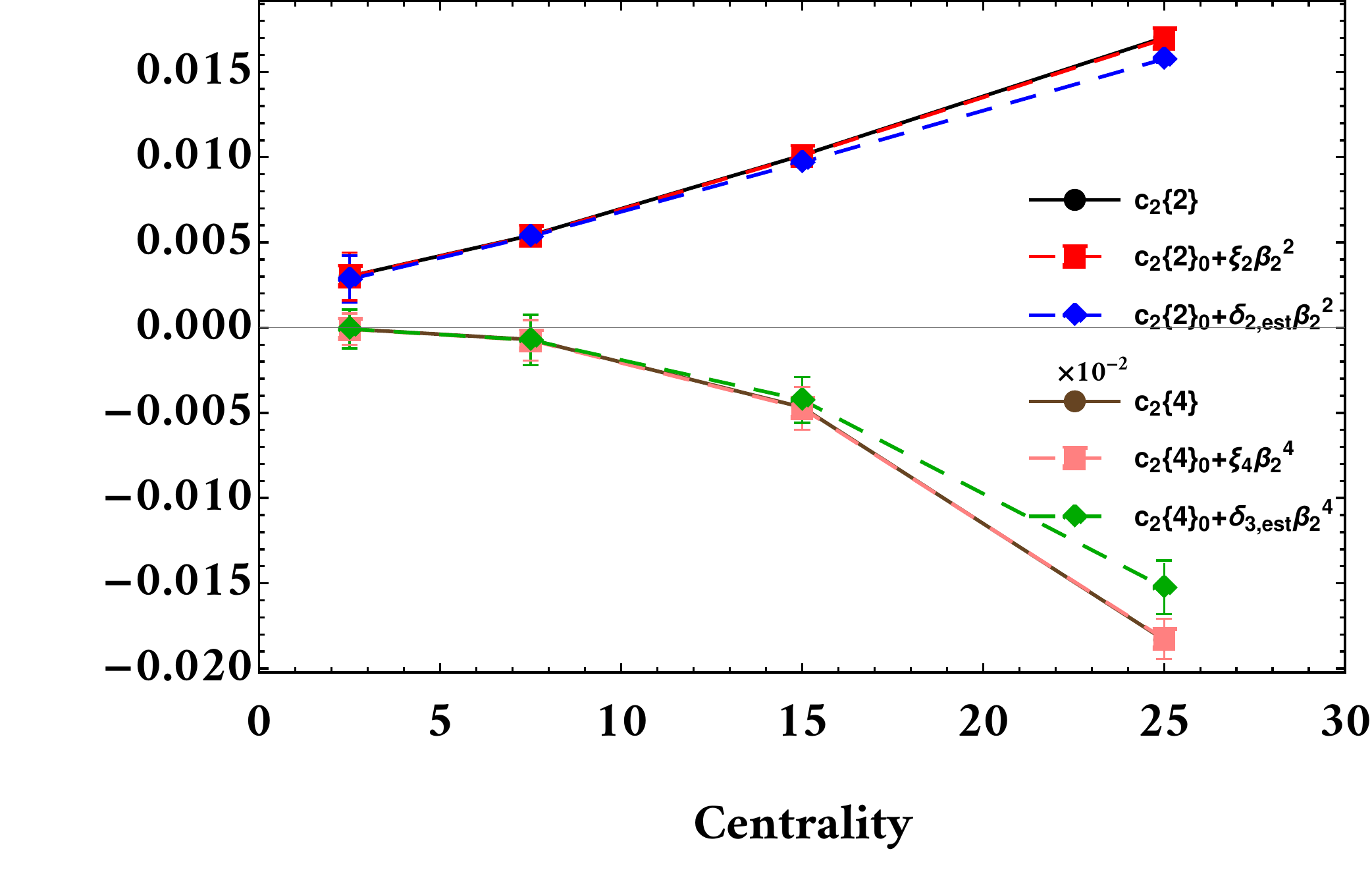}
	\end{tabular}		
	\caption{(Color online) Comparison of $c_2\{2\}$ and $c_2\{4\}$ obtained from DD data with their estimates using SS. The value of $c_2\{4\}$ was multiplied by $100$.} 
	\label{fig11}
\end{figure}		
Now, we obtain $\xi_{2k,m}$ by equating Eqs.~\ref{qq10} and \ref{qxi}. Since we are interested in $\beta_{2}$ terms, we seek an expression for $\xi_{2k,2}$ as a function of $\delta$ and $\beta_{2}$. This is given by:
\begin{equation}\label{qqA16}
\begin{split}
\xi_2\equiv\xi_{2,2}^2&=\delta_{1,2}^2+\delta_{2,2}^2+2\frac{\delta_{1,2}\bar{v}_{2,0}}{\beta_2},\\
\xi_4\equiv\xi_{4,2}^4&=-\delta_{1,2}^4+\delta_{3,2}^4-4\frac{\delta_{1,2}^3\bar{v}_{2,0}}{\beta_2}\\&-6\frac{\delta_{1,2}^2\bar{v}_{2,0}^2}{\beta_2^2}-4\frac{\delta_{1,2}\bar{v}_{2,0}^3}{\beta_2^3}.
\end{split}	
\end{equation}
Plugging $\delta_{1,est}=\delta_{1,2}$, $\delta_{2,est}=\delta_{2,2}^2$, and $\delta_{3,est}=\delta_{3,2}^4$ in Eq.~\ref{qqA16} we arrive at:
\begin{equation}
\begin{split}
\xi_2\equiv\xi_{2,2}^2&=\delta_{1,est}^2+\delta_{2,est}+2\frac{\delta_{1,est}\bar{v}_{2,0}}{\beta_2},\\
\xi_4\equiv\xi_{4,2}^4&=-\delta_{1,est}^4+\delta_{3,est}-4\frac{\delta_{1,est}^3\bar{v}_{2,0}}{\beta_2}\\&-6\frac{\delta_{1,est}^2\bar{v}_{2,0}^2}{\beta_2^2}-4\frac{\delta_{1,2}\bar{v}_{2,0}^3}{\beta_2^3}.
\end{split}	
\end{equation}
For the particular values listed in Table.~\ref{tab1}, the coefficients $\xi_{2}$ and $\xi_{4}$ are found. 
\begin{table}[t!]
	\caption{The estimated coefficients in Eq.~\ref{qq10} are shown at different centralities.}
	\begin{supertabular}{| c | c | c | c |}
		\hline
		$\%$& $\delta_{1,est}$ & $\delta_{2,est}$ & $\delta_{3,est}$  \\ \hline
		$0-5$ & $0.014\pm0.0082$ & $0.020\pm0.0014$ & $-0.0002\pm0.00002$ \\ \hline
		$5-10$ & $0.001\pm0.0001$ & $0.022\pm0.0004$ & $-0.0004\pm0.00003$  \\ \hline
		$10-20$ & $0.0088\pm0.0006$ & $0.017\pm0.0003$ & $-0.0004\pm0.00003$  \\	\hline
		$20-30$ & $0.020\pm0.0006$ & $0.004\pm0.0002$ & $0.0015\pm0.00003$ \\ \hline
	\end{supertabular}
	\label{tab1}
\end{table}
We plot the $c_{2}\{2\}$ and $c_{2}\{4\}$ in Fig.~\ref{fig11}. In this plot, the solid black and brown represent the true centrality dependence of the aforementioned quantities. Moreover, the dashed red and pink lines are derived from our estimation. There is a good agreement between the true and estimated values. Moreover, this figure shows that if we consider $\xi_2\approx\delta_{2,est}$ (blue dashed line) and  $\xi_4\approx\delta_{3,est}$ (green dashed line), we can find a reasonable approximation for them from $0$ to $20\%$ centralities. Plugging Eq.~\ref{qcn} into \ref{qq10} and using the approximation described, we obtain:
\begin{equation}
R_2\{2k\}_{D}-R_2\{2k\}_{0}=c_2\{2k\}_{D}-c_2\{2k\}_{0}.
\end{equation}

This leads to the same value of averaged ellipticity for both SS and DD collisions, i.e. $\bar{v}_D\approx\bar{v}_S$ (see Sec.~\ref{sec6}). While the studied nuclei have an mass number near one another, it is possible to access this information.
To compare the observables of DD with those of SS, we study the ratio of $2k$-particle correlation functions:
\begin{equation}\label{q19}
\begin{split}
\frac{c_2\{2\}_{D}}{c_2\{2\}_0}&=1+\frac{\xi_{2}\beta_2^{2}}{c_2\{2\}_0},\\
\frac{c_2\{4\}_{D}}{c_2\{4\}_0}&=1+\frac{\xi_{4}\beta_2^{4}}{c_2\{4\}_0}.
\end{split}
\end{equation}
Keep in mind that we have only turned on the quadrupole deformation in Eq.~\ref{qcn}. Using the generated data for both SS (i.e., PbPb) and DD (i.e., UU), we obtain $c_2\{2\}_{D}/c_2\{2\}_0\approx2$ and $c_2\{4\}_{D}/c_2\{4\}_0\approx-5$ at most central collisions. The values imply that we have $c_2\{2\}_0\approx\xi_{2}\beta_2^{2}$ and $c_2\{4\}_0\approx(-1/6)\xi_{4}\beta_2^{4}$. 
The effect of deformation on 2 and 4-particle correlation functions is significant and cannot be ignored.       

\section{Ellipticity}\label{sec6} 
One of the main results from studying flow harmonics is that the averaged ellipticity $\bar{v}_{2n}$ is non-zero. This leads us to look for an accessible estimation of $\bar{v}_{2n}$ experimentally. Since we are interested in $v_2$ for SS and DD collisions, we present a possible approach to observe this quantity. Let us start with a 2D distribution of $(v_{2,x},v_{2,y})$ in Fig.~\ref{fig9}. As it turns out, there is a non-vanishing $\bar{v}_{2}$ for collisions of deformed as well as spherical nuclei. Despite the large size of $(v_{2,x},v_{2,y})$ distributions for DD collisions, the values of averaged ellipticity are the same.
However, the path to find the estimations of $\bar{v}_2$ for SS and DD collisions is different. 
\begin{figure}[t!]
	\begin{tabular}{c}
		\includegraphics[scale=0.42]{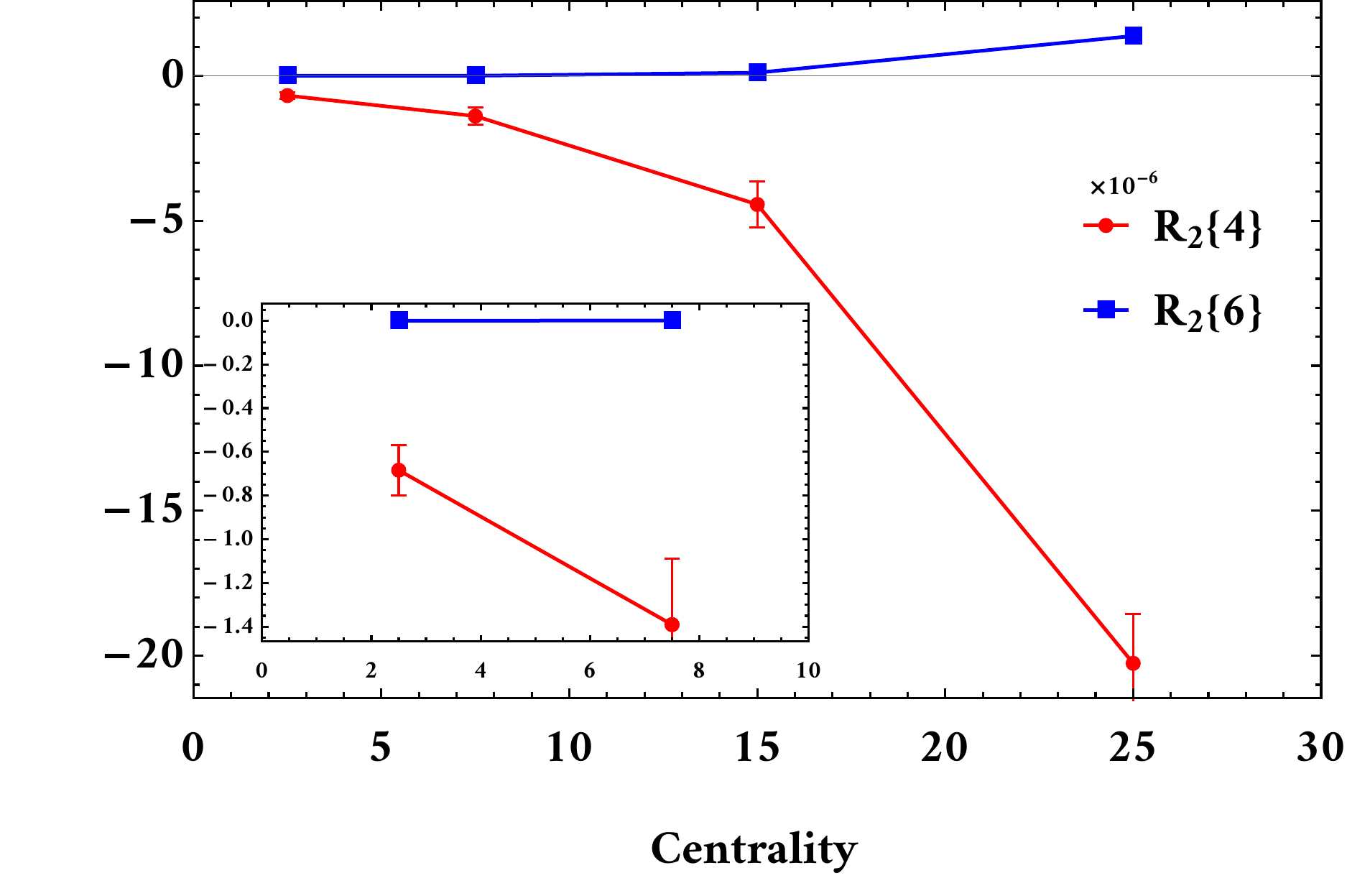}\\
		\includegraphics[scale=0.42]{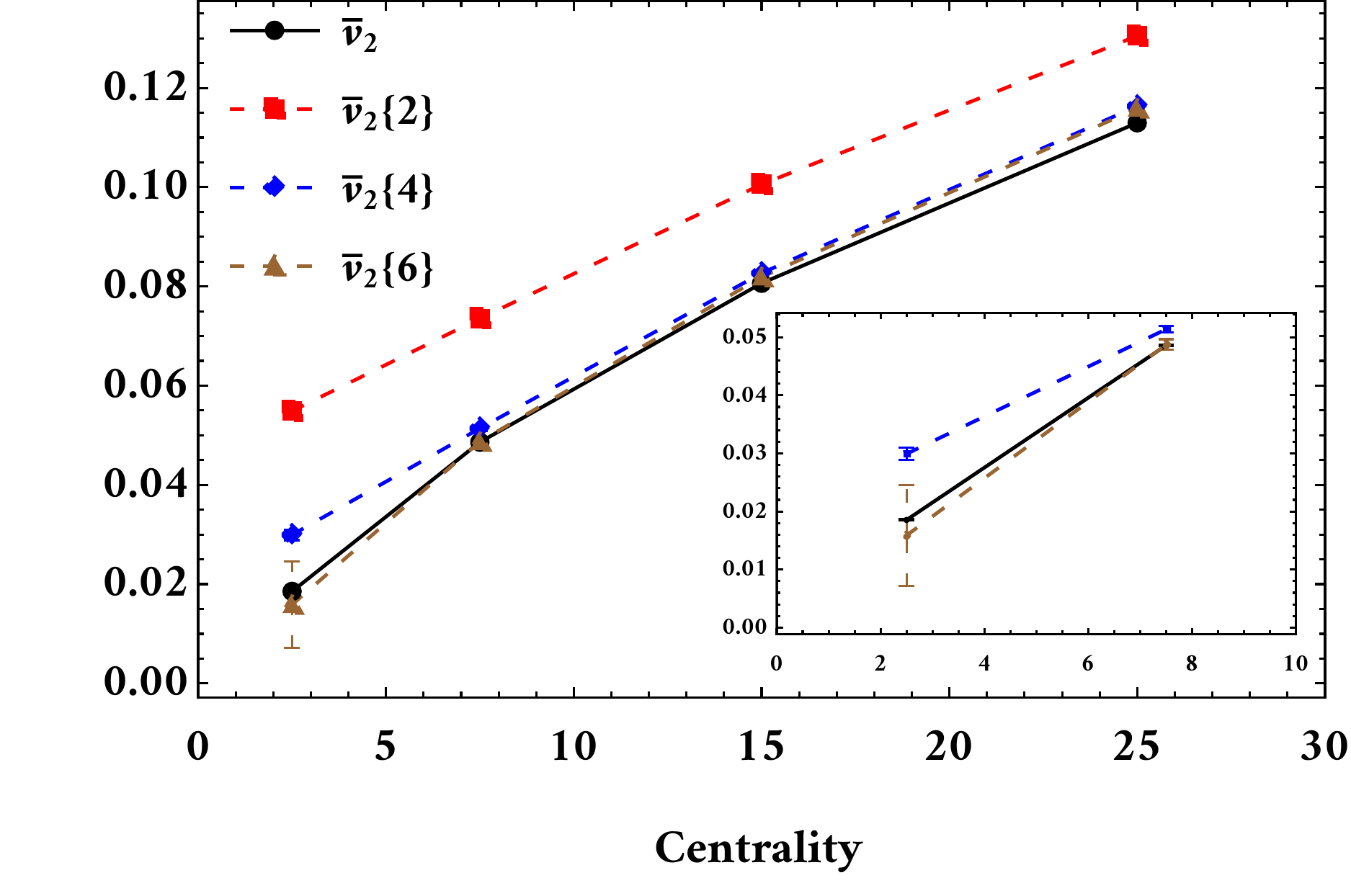}
	\end{tabular}		
	\caption{(Color online) Here we show a comparison of $R_2\{4\}$ with $R_2\{6\}$ in the top panel and the estimated values of $\bar{v}_{2}$ for different collisions as a function of centrality.} 
	\label{fig10}
\end{figure} 
At first, we prefer to present this estimation for deformed nuclei collisions. To do this, we start with the closest estimate of distribution $P_r(v_2)$ for DD collisions which is given by $BG+P_2(v_2)$. This means the higher order correction terms, i.e. $R_2\{6\}$, are very small such that $R_2\{6\}\approx0$. To verify this, we plotted this cumulant in Fig.~\ref{fig10}. As demonstrated, the magnitude of $R_2\{4\}$ for various centralities is larger than $R_2\{6\}$. Moreover, at $0-5\%$ and $5-10\%$ centalities $R_2\{6\}$ is closer to zero. Therefore, we estimate the value of $R_2\{2k\}$ for $k=1,2,3$ by considering:            	
\begin{equation}\label{qq13} 
\begin{split}
R_2\{2\}&=c_2\{2\}-\bar{v}_2^2\approx 0 \;\implies\; \bar{v}_2\{2\}\approx\left(c_2\{2\}\right)^{1/2},\\
&\hspace*{3cm}\text{or}\\
R_2\{4\}&=c_2\{4\}+\bar{v}_2^4\approx 0 \;\implies\; \bar{v}_2\{4\}\approx\left(-c_2\{4\}\right)^{1/4},\\
&\hspace*{3cm}\text{or}\\
R_2\{6\}&=c_2\{6\}-4\bar{v}_2^6\approx 0 \;\implies\; \bar{v}_2\{6\}\approx\left(c_2\{6\}/4\right)^{1/6}. 
\end{split}
\end{equation}
Focusing on the first condition, we find that in this case all the $\gamma_{2k}$ in Eq.~\ref{qqq8} diverge unless $R_2\{2k\}=0$. This leads to finding a delta function for $P(v_{2,x},v_{2,y})$, thus it is not compatible with the experimental observation. As the bottom panel in Fig.~\ref{fig10} depicts, $\bar{v}_2\{2\}$ is not a suitable candidate of $\bar{v}_2$. Having a Bessel-Gaussian distribution is the result of choosing the second line \cite{Jia:2022qgl}. This implies the behavior of SS and DD distributions is similar and we see no effect of nuclei deformity using distribution analysis. This is in contrast to our conclusion so far. The mini panel in the bottom plot in Fig.~\ref{fig10} indicates this estimation is not accurate at most central collisions. Of course, $\bar{v}_2\{4\}$ is a suitable choice to estimate averaged ellipticity at large centralities.
Finally, we arrive at the last line of Eq.~\ref{qq13}. This implies a truncation at $k=2$. This is in agreement with our results in sec.~\ref{sec3}. To conclude this section, the closest estimate of $\bar{v}_2$ is given by $\bar{v}_2\{6\}$. In contrast to $\bar{v}_2\{4\}$, only $\bar{v}_2\{6\}$ explains $\bar{v}_2$ at most central collisions where the maximum deformity is expected to be observed. Since PbPb data can be explained by BG distribution, we find that $\bar{v}_{2,S}=\bar{v}_{2,S}\{4\}$. Concerning this argument and derived relation $\bar{v}_D\approx\bar{v}_S$ in the Sec.~\ref{sec5}, one can find $\bar{v}_{2,D}=\bar{v}_{2,D}\{6\}=\bar{v}_{2,S}\{4\}$ as well. This means that we can determine the averaged ellipticity of DD collisions with the observables of SS ones.
\\
\vspace*{.25cm}
\section{Conclusions}\label{sec.con}
Motivated by the collisions of deformed nuclei, in this paper, we studied the flow distribution of symmetric and deformed nuclei. In the first part of this manuscript, we presented a systematic approach to calculating the corresponding cumulants for SS and DD collisions. It was shown that in most central collisions there is no difference between different nuclei for $c_2\{2k\}|_{k>1}$. 
To be able to distinguish between different ions, we considered the effect of the shift parameter $\bar v_{n}$. Then, we scrutinized the effect of a different form of deformation, including quadrupole $\beta_{2}$ as well as octupole $\beta_{3}$, through the shifted cumulants. We observed that the shift parameter manifest the differences between the cumulants of deformed and spherical nuclei clearly. 
\par
Using the obtained information from cumulative studies, we calculated the corresponding distribution of flow harmonics. It was shown that, after keeping an appropriate number of terms, the resulting distribution described the data very well. Comparing the distribution of deformed and spherical nuclei reveals the effect of various kinds of deformation on flow harmonics. As it turns out, increasing the quadrupole magnitude $\beta_{2}$, deformation results in a broader distribution compared to the symmetric one, see for example Fig.~\ref{fig7}. We, further, discussed the possibility of interpolating from spherical to deformed nuclei by including appropriate corrections. We examined this idea where we could generate the deformed correlation as well as cumulants with high precision.
\par
Finally, we discussed a possible way to measure the shift parameter through the analysis of different radial cumulants for deformed nuclei in central collisions. We observed for asymmetric nuclei the most appropriate choice is the measurement of $\bar{v}_{2}\{6\}$. It would be interesting to extend this work to the collision of $\text{Ru}$-$\text{Ru}$ and $\text{Zr}$-$\text{Zr}$ using full hydrodynamic simulations which are more relevant for the isobar program. The aim of such studies is to extract the effect related to CME from the background. We postpone these subjects to future studies.

\section*{Acknowledgment}
We thank Jiangyong Jia and Giuliano Giacalone for their useful comments.
We are thankful to Wilke van der Schee for helpful discussions and
invaluable feedback. H.M.~thanks CERN-TH group for the support. H.M.~is funded by the Cluster of Excellence {\em Precision Physics, Fundamental Interactions, and Structure of Matter} (PRISMA$^+$ EXC 2118/1) funded by the German Research Foundation (DFG) within the German Excellence Strategy (Project ID 39083149).

\end{document}